\documentclass[prb,twocolumn,showpacs,superscriptaddress,amsmath,floatfix]{revtex4}
\usepackage{graphicx}
\usepackage{amssymb}
\usepackage{multirow}
\usepackage{longtable}
\usepackage{booktabs}
\begin{document}

\title{The Valence Transition Model of Pseudogap, Charge-Order and Superconductivity in Electron- and Hole-Doped Copper Oxides.}
\author{Sumit Mazumdar}
\affiliation{Department of Physics, University of Arizona
Tucson, AZ 85721}
\affiliation{Department of Chemistry and Biochemistry, University of Arizona, Tucson, AZ 85721}
\affiliation{College of Optical Sciences, University of Arizona, Tucson, AZ 85721}
\date{\today}
\begin{abstract}
We present a valence transition model for electron- and hole-doped cuprates, 
within which there occurs a 
discrete jump in ionicity Cu$^{2+} \to$ Cu$^{1+}$ in both families upon doping, at or near optimal doping in the conventionally
prepared electron-doped compounds and at the pseudogap phase transition in the hole-doped materials. In thin films of the T$^\prime$
compounds, the valence transition has occurred already in the undoped state. 
The phenomenology of the  valence transition is closely related to that of the neutral-to-ionic transition in mixed-stack organic charge-transfer
solids. Doped cuprates have negative charge-transfer gaps, just as
rare earth nickelates and BaBiO$_3$. The unusually high ionization energy of the closed shell Cu$^{1+}$ ion,
taken together with the doping-driven reduction in three-dimensional Madelung energy and gain in two-dimensional delocalization 
energy in the negative charge transfer gap state drives the transition in the cuprates. 
The combined effects of strong correlations and
small $d-p$ electron hoppings ensure that the systems behave as effective $\frac{1}{2}$-filled Cu-band with the closed shell electronically inactive
O$^{2-}$ ions in the undoped state, and as correlated two-dimensional geometrically frustrated $\frac{1}{4}$-filled oxygen
hole-band, now
with electronically inactive closed-shell Cu$^{1+}$ ions, in the doped state. The model thus gives microscopic justification for the two-fluid models
suggested by many authors. The theory gives the simplest yet most comprehensive understanding of experiments in the normal states.  
The robust commensurate antiferromagnetism in the conventional T$^\prime$ crystals, the strong role
of oxygen deficiency in driving superconductivity and charge carrier sign corresponding to holes at optimal doping are all manifestations of the
same quantum state. In the hole-doped pseudogapped state, there occurs a biaxial commensurate period 4 charge density wave state consisting of
O$^{1-}$-Cu$^{1+}$-O$^{1-}$ spin-singlets, that coexists with broken rotational C$_4$ symmetry due to intraunit cell oxygen inequivalence. 
Finite domains of this broken symmetry state will exhibit two-dimensional chirality and the polar Kerr effect.
Superconductivity within the model results from a destabilization of the $\frac{1}{4}$-filled band paired Wigner crystal
[Phys. Rev. B {\bf 93}, 165110 and  {\bf 93}, 205111]. We posit that a similar valence transition,
Ir$^{4+} \to$ Ir$^{3+}$, occurs upon electron doping SrIr$_2$O$_4$. We make testable experimental predictions on cuprates 
including superoxygenated La$_2$CuO$_{4+\delta}$
and iridates. Finally, as indirect evidence for the valence bond theory of superconductivity proposed here, 
we note that there exist an unusually large number of unconventional superconductors 
that exhibit superconductivity proximate to exotic charge ordered states, whose bandfillings are universally $\frac{1}{4}$ or
$\frac{3}{4}$, exactly where the paired Wigner crystal is most stable. 

\end{abstract}
\maketitle
\section {\bf Introduction.}
\label{intro}
The phenomenon of high temperature superconductivity (SC) in layered copper oxides has now been known for more than 
three decades \cite{Bednorz86a}. In spite of intense
experimental and theoretical studies, correlated-electron SC
continues to be a formidable problem. Not only is there no consensus among theorists over the mechanism of 
SC itself,
the nature of even the ``normal'' state of the weakly doped parent semiconducting materials continues to be mysterious.
The pseudogap (PG) state in the underdoped hole-based cuprates has been intensively studied by experimentalists
and theorists alike; it is, however, probably fair to say that new experimental revelations on cleaner samples using sophisticated
experimental techniques have served mostly to indicate the shortcomings of theoretical approaches. The most well known example of this is the
apparent contradiction between experiments that suggest that the origin of PG is due to fluctuating SC
\cite{Norman98a,Wang05a,Wang06a,Li10b,Kanigel06a,Chatterjee11a,Dubroka11a,Mishra14a}
versus more recent ones that have indicated the existence of a charge ordered (CO) phase within the PG 
\cite{Hoffman02a,Howald03a,Hanaguri04a,Shen05a,Blanco-Canosa13a,SilvaNeto14a,Comin14a,Tabis14a,Hashimoto14a,Wu11a,Wu13a,Wu15a,Cai16a,Mesaros16a}. To obviate the 
preformed pairs versus competing order conundrum some investigators have proposed that the CO is a density wave of Cooper pairs 
\cite{Anderson04b,Franz04a,Tesanovic04a,Chen04a,Vojta08a,Hamidian16a,Cai16a,Mesaros16a}. Neither the nature of this density wave, nor the
mechanism of its formation is understood. 

In the present paper we present an evidence-based phenomenological theory of cuprates that is substantively different from 
all existing approaches. First presented very early on \cite{Mazumdar89a,Mazumdar89b}, the theory treats cuprates as correlated charge-transfer
semiconductors \cite{Zaanen85a} that can undergo a phase transition absent in Mott-Hubbard semiconductors. 
The fundamental principle behind the
Zaanen-Sawatzky-Allen (ZSA) classification scheme of transition metal compounds as  charge-transfer semiconductors versus Mott-Hubbard 
semiconductors \cite{Zaanen85a} is well understood. The valence transition model goes beyond the ZSA theory by positing 
that charge-transfer semiconductors with the metal cation with electron configuration M$^{n+}$ are susceptible
to valence transition M$^{n+}$ $\to$ M$^{(n-1)+}$, if the latter has a closed-shell configuration and is hence particularly stable.
Such a {\it discrete} jump in ionicity is different from both 
mixed valence or covalency involving the cation and the anion, and cannot be captured within band or density functional theories (DFT). 
Closely related theoretical models of first order transitions between nearly integer valences have been studied over 
several decades in the  context of temperature, pressure, and light-induced neutral-to-ionic transitions 
in strongly correlated organic mixed-stack charge-transfer solids
\cite{Torrance81a,Torrance81b,Tokura89a,Kosihara90a,Horiuchi03a,Mazumdar78a,Pati01a,Nagaosa86a}.  
Also closely related is the current idea of negative charge-transfer gap \cite{Mizokawa91a,Khomskii01a} 
in rare-earth nickelates (RE)NiO$_3$, where with the 
exception of LaNiO$_3$, the true electron configuration of the transition metal is Ni$^{2+}$ ($d^8$) instead of the Ni$^{3+}$ ($d^7$)
expected from formal charge counting \cite{Torrance92a,Green16a,Bisogni16a,Shamblin18a}. 
Note that the true metal-ion electron configuration requires that fully one-third of the oxygens are O$^{1-}$. The predicted high concentration of
O$^{1-}$ has been experimentally confirmed \cite{Bisogni16a,Shamblin18a}. 
Interestingly, an early theory of the metal-insulator transition \cite{Torrance92a} in (RE)NiO$_3$ is very closely related to the neutral-ionic
transition model. Finally, lower than formal charge has also been recognized recently in the perovskite BaBiO$_3$, where
again the true electron configuration of the Bi-ions is a homogenous Bi$^{3+}$, as opposed to what was believed before, alternating
Bi$^{3+}$ and  Bi$^{5+}$ ions \cite{Plumb16a,Khazraie18a}.
We will not only be interested in when negative charge-transfer gap is most likely, but whether there can occur a real transition from positive to negative 
charge-transfer gap.

The fundamental reason behind this negative
charge-transfer gap is that high positive charges on cations are intrinsically unstable.
It then stands to reason that in systems where the charge-transfer gap is small to begin with, there is the likelihood
of a dopant-induced valence transition \cite{Mazumdar89a,Mazumdar89b,Khomskii01a}.
We argue here that such a transition indeed occurs in the cuprates (and
also in some other systems that are being treated as Mott-Hubbard semiconductors, for e.g., nominally Ir$^{4+}$ compounds, see below).
Although the valence transition to the PG state in the cuprates (as well as homogenous population of Bi$^{3+}$ in doped BaBiO$_3$)
was actually {\it predicted} within the valence transition model \cite{Mazumdar89b},
theoretical work along this direction was not
continued, as there appeared to be no obvious explanation of SC within the theory. The motivation for the present work
comes from, (a) the recent demonstration of the enhancement of superconducting pair-pair correlations by Coulomb
interactions within the theoretical model appropriate for the post-valence transition cuprates \cite{Gomes16a,DeSilva16a,Clay18a},
and (b) experimental discoveries (see section \ref{challenges}) over the intervening decades
that strongly justify the theory. The goal of the present paper is to demonstrate that the
model gives simultaneously the simplest yet most comprehensive explanations of the peculiar features of the cuprate families, both electron-
and hole-doped. SC in both families (and in several other correlated-electron superconductors, see Appendix)
can be understood within a valence
bond (VB) theory \cite{Gomes16a,DeSilva16a,Clay18a} that is influenced by Anderson's resonating valence bond (RVB) \cite{Anderson87a} theory but 
is substantially different. Indeed, our theory is at the interface of the RVB theory and the oldest version of the so-called bipolaron
theory of superconductivity \cite{Chakraverty85a}, that can be thought of as a precursor to the RVB theory. The spin-singlet bonds
anticipated in the superconducting states of the materials discussed in reference \onlinecite{Chakraverty85a} can arise from electron-electron interactions 
\cite{Gomes16a,DeSilva16a}, as opposed to being driven by overly strong electron-phonon couplings, as assumed in the earlier literature \cite{Chakraverty85a}.

We begin this work by presenting  in section ~\ref{challenges} a list of experiments that we believe any minimal model of the normal state of the real materials
should be able to explain. Such a listing is essential to elucidate the full scope of the theoretical challenge, especially in the
context of the present theoretical approach which challenges the widely accepted notion that superconducting cuprates can be described within the
weakly doped Mott-Hubbard semiconductor model. Each experimental observation listed has been
considered individually crucial for understanding cuprate physics by multiple research groups.
However, there exists no theoretical work that has attempted to explain the entire list.   
It is only when these experimental observations are considered together that the need for a theoretical model substantively
different from any of the  ``traditional'' or ``accepted'' models becomes obvious.

Following the presentation of this list of experimental challenges, in section ~\ref{model} we present the
theory of valence transition, as applied to cuprates. In section ~\ref{explanations} we revisit the experiments of section ~\ref{challenges} to
show how all of these observations actually are to be expected within the valence transition model: all the supposedly exotic phases, in
both electron- and hole-doped materials are manifestations of the same quantum state. We then present a VB theory of SC, partial numerical
evidence for which has been presented recently \cite{Gomes16a,DeSilva16a}.
We believe that the valence transition model is applicable to other transition metal oxides where also the lower ionic charge corresponds
to closed shell. In section ~\ref{iridates} we discuss recent experimental observations of a PG state in doped SrIr$_2$O$_4$, that we believe 
can be understood
within a Ir$^{4+} \to$ Ir$^{3+}$ valence transition scenario.
In section ~\ref{conclusions} we present our conclusions, and also make a series of
testable experimental predictions uniquely specific to the present theory.
A basic contention of the present work is that correlated-electron SC can and will result in geometrically frustrated $\frac{1}{4}$-filled 
(or $\frac{3}{4}$-filled) systems. In order to avoid confusion, henceforth we will mostly refer to carrier density $\rho$, which is more
appropriate for correlated-electron systems than ``band-filling'', and which is 0.5 at both these fillings. 
In the Appendix we list several different families of correlated-electron superconductors that in all cases have $\rho=0.5$.

\section{Theoretical Challenges}
\label{challenges}

We begin with the electron-doped cuprates and follow up with the hole-doped materials.
Recent experimental discoveries with electron-doped cuprates challenge  more strongly than in the hole superconductors
the notion that the antiferromagnetism (AFM) in the parent semiconductor drives SC.
Further, there exist significantly less theoretical work on the former.

\subsection{Experimental puzzles: electron-doped materials}
\label{electrons}
Several recent reviews have given excellent discussions of the experimental developments on the electron-doped cuprates
\cite{Armitage10a,Naito16a,Adachi17a}.
There is general agreement
that the difference between the electron- and hole-doped cuprates originates from their different crystal structures ($T^\prime$ in the
former versus predominantly $T$ in the latter). Developing a theory for the electron-doped systems has been difficult, as some of the
experimental observations are universally shared between electron and hole-doped materials, while others are unique to one or the
other family.
The situation has become more confusing in recent years, as with the development of specialized reduction annealing processes
the boundary between the AFM and the superconducting phase has shifted to smaller and smaller doping concentrations, and even the
completely undoped $T^\prime$ compounds have been found to be superconducting \cite{Naito16a}. In the following we make distinction
between ``conventionally annealed'' and ``specially annealed''
$T^\prime$ compounds.
\vskip 0.5pc
\noindent \underbar{(i) Robust AFM in the conventional T$^\prime$ compounds.} This is the feature of the electron-doped materials that has attracted the most
attention. In both Nd$_{2-x}$Ce$_x$CuO$_{4-\delta}$ (NCCO) and Pr$_{2-x}$Ce$_x$CuO$_{4-\delta}$ (PCCO) {\it commensurate} AFM persists upto doping $x \sim 0.13-0.14$ and SC
occurs over the narrow doping concentration \cite{Armitage10a} $x=0.15-0.18$. Dynamical mean field theory (DMFT) calculations have
ascribed this to the undoped $T^\prime$ compounds being ``weakly correlated Slater antiferromagnets'', as opposed to the
$T$ compounds that are more strongly correlated Mott-Hubbard semiconductors within the calculations \cite{Das09a,Weber10a}.
While it is self-evident that the absence of apical oxygens in the $T^\prime$
structure gives a smaller Madelung energy stabilization of the highly ionic Cu$^{2+}$-based description of the parent
semiconductor \cite{Torrance89a},
it is not intuitively apparent why the Hubbard $U$ should be smaller in the $T^\prime$ compounds as a consequence. More importantly, it is
unlikely that the experimental peculiarities listed below can be understood within this picture.
\vskip 0.5pc
\noindent \underbar{(ii) Absence of coexisting SC and AFM in conventional}
\noindent \underbar{ $T^\prime$ compounds.}
Experiments by several research groups have indicated that SC and AFM do
not coexist.
A quantum critical point at $x \simeq 0.13$ separating AFM and SC has been claimed from inelastic
magnetic neutron-scattering measurements in NCCO \cite{Motoyama07a}. Even though SC in LCCO appears at significantly smaller doping concentration,
muon spin rotation measurements have found a similar phase boundary \cite{Saadaoui15a} between three-dimensional (3D) static AFM and SC
(the authors do not preclude fluctuating two-dimensional (2D) magnetic order).
Magnetic field-induced quantum phase transition from the superconducting state to a {\it commensurate} AFM state has
been found in NCCO \cite{Kang03a}.
\vskip 0.5pc
\noindent \underbar{(iii) RE size dependence of AFM-SC boundary and T$_c$.}
The doping concentration range over which SC is observed in the family (RE)$_{1-x}$Ce$_x$CuO$_{4-\delta}$ and the superconducting T$_c$ both
increase dramatically with the size of the RE ion \cite{Naito02a} (see for example, Fig.~5 in reference \onlinecite{Naito16a}).
The doping range over which La$_{2-x}$Ce$_x$CuO$_{4-\delta}$ (LCCO) with the very large La$^{3+}$ ion is a superconductor ($x \geq 0.08$) as well as
its superconducting T$_c$ are significantly
its superconducting T$_c$ are significantly
larger \cite{Saadaoui15a} than those in NCCO and PCCO.
\vskip 0.5pc
\noindent \underbar{(iv) Oxygen deficiency as a requirement for SC.} A characteristic of the electron-doped cuprates that has received
very strong interest from experimentalists (and in contrast, very little interest from theorists) is that for SC to occur it is absolutely essential that there is some reduction of oxygen content
({\it i.e.}, $\delta \neq 0$) \cite{Armitage10a,Naito16a}. It is accepted that this is {\it not} due to self-doping, in view of the following
observations, (a) the
deficiency that is required is very small ($\delta \leq 0.04$), and (b) it is not possible to compensate for the
lack of deficiency by addition of extra Ce \cite{Kim93a,Navarro01a,Armitage10a}.
In one of the most intriguing experiments, single crystals of NCCO annealed in small oxygen partial
pressures $p_{O_2}$ at different temperatures T showed two distinct regimes, with higher $p_{O_2}$ leading to nonsuperconducting materials and lower $p_{O_2}$
to superconductors. The boundary between these two regimes coincides with the phase stability line between CuO (oxidation state Cu$^{2+}$)
and Cu$_2$O
(oxidation state Cu$^{1+}$), {\it with the superconducting electron-doped cuprates lying firmly in the region corresponding to Cu$_2$O}
\cite{Kim93a,Navarro01a}.

The above observations are similar to what had been observed in early experiments with
the fluorine-doped electron superconductor Nd$_2$CuO$_{4-x}$F$_x$, which is superconducting \cite{James89a} for $x=0.4$.
The material remains semiconducting when annealed at high temperatures in air, but is superconducting when annealed in nitrogen.
While this compound has been far less studied than other compounds, there are other similarities between this system and the more usual
electron-doped superconductors.
\vskip 0.5pc
\noindent \underbar{(v) SC in ``underdoped'' and undoped $T^\prime$ materials.} With the discoveries of specialized annealing techniques SC has
been found at lower and lower dopings, at Ce-concentration $x \simeq 0.04$ and $x \simeq 0.05$, respectively, in PCCO \cite{Brinkmann95a} and
Pr$_{1.3-x}$La$_{0.7}$Ce$_x$CuO$_4$ (PLCCO) \cite{Adachi17a}. Using metal-organic decomposition Naito and coworkers have obtained
SC in {\it undoped} $T^\prime$-(RE)$_2$CuO$_4$ with R = Pr, Nd, Sm, Eu and Gd (Gd$_2$CuO$_4$ with the smallest RE ionic radius
does not exhibit SC in the bulk) \cite{Naito16a}. T$_c$ in these unconventional underdoped and undoped materials are higher than
that in the conventionally doped systems and is maximum for zero Ce-doping \cite{Naito16a}.
Naito {\it et al.}
have demonstrated quite clearly that the condition for reaching SC is removal of excess apical impurity oxygens,
casting severe doubt on the conventional wisdom that SC is a consequence of doping a Mott-Hubbard semiconductor.
Importantly, removal of excess apical oxygens renders the Cu sites nonmagnetic, in agreement with the observation that SC and AFM are
noncoexisting \cite{Saadaoui15a,Motoyama07a}.
As in the earlier annealing experiments \cite{Kim93a,Navarro01a}, the authors found that both $T$ and $T^\prime$ cuprates lie significantly below the stability
line of CuO and close to that of Cu$_2$O in the $p_{O_2}$-1/T plane.
\vskip 0.5pc
\noindent \underbar{(vi) Carrier concentration different from dopant}
\noindent \underbar{concentration.} The actual effective carrier concentration in the electron-doped compounds has always been
a mystery, given the persistence of AFM upto large $x$ in the conventional materials. The successful synthesis of undoped $T^\prime$ superconductors
has brought this question to the fore.
A number of recent experimental investigations \cite{Horio16a,Wei16a,Song17a} have confirmed that
reduction annealing by removing apical oxygens severely reduces the stability of the AFM phase and {\it introduces additional carriers
by some mechanism
that is as yet not understood.} The actual carrier density even in conventional materials is different from what would be guessed from the Ce concentration
alone \cite{Horio16a,Song17a}. Horio {\it et al.}, in particular, find complete absence of AFM and a Fermi surface much larger than
expected in $x=0.1$ PLCCO from
angle-dependent photoemission spectroscopy (ARPES) measurment \cite{Horio16a}.

\vskip 0.5pc
\noindent \underbar{(vii) Sign of the charge carrier.} Hall coefficient measurements in the conventional $T^\prime$ materials have found R$_H$ that is
negative at small $x$, but that then increases with increasing $x$ and becomes positive in the overdoped
region immediately beyond the dopant concentration range where SC is seen \cite{Dagan04a,Gauthier07a,Krockenberger13a}. These results agree with earlier
ARPES studies that found large holelike Fermi surface \cite{Armitage02a} in NCCO for $x>0.1$.
More perplexing are the results of similar measurements in samples obtained with specialized annealing, where
positive R$_H$ is found for the undoped superconductors \cite{Adachi17a,Krockenberger13a}. Various phenomenological
two-band models have been proposed to explain this
unexpected carrier sign. In particular, it has been proposed that the undoped materials without apical oxygens are already metals with the charge
carriers coming from both Cu and O. We will provide an alternate explanation in better agreement with other observations.
\vskip 0.5pc
\noindent \underbar{(viii) Cu NMR and NQR.} Large reduction in $^{63,65}$Cu NMR intensity at low temperatures and optimal doping is a characteristic
of electron-doped cuprates that is also not understood. Unexpectedly small NQR frequency is found in the normal states of optimally doped
electron-doped cuprates \cite{Abe89a,Williams07a,Wu14a} as compared to the NQR frequencies in the parent semiconductors. The ultrasmall NQR
frequencies correspond to tiny electric field gradient (EFG) that is surprising within the standard picture of doping that would leave the majority
of the Cu-ions as Cu$^{2+}$ with 3d$^9$ configuration. The earliest work \cite{Abe89a} had therefore suggested that {\it there are dramatic differences
in the electronic environments about the Cu-sites in the weakly versus optimally doped materials}, a conclusion that the valence transition
model justifies.

\vskip 0.5pc
\noindent \underbar{(ix) Charge-order (CO)}. CO has now been found in nearly all hole-doped compounds and is discussed in greater detail in the next subsection.
While many different mechanisms have been proposed for the formation of a CO phase, in the hole-doped materials it has become clear that nesting-based
scenarios do not explain
the CO (see below). Assuming the same is true for CO in NCCO \cite{SilvaNeto15a,SilvaNeto16a} and LCCO \cite{SilvaNeto16a}, theoretical explanation of
CO in the electron-doped materials faces even greater difficulty. CO periodicities of [0.23 $\pm$ 0.04]Q and [0.24 $\pm$ 0.04]Q (Q=2$\pi/a_0$,
where $a_0$ is the Cu-O-Cu lattice constant) at the optimal
doping concentrations of 0.14 and 0.15 pose particular challenge, in view of of their being so close to the
doping-independent commensurate periodicity 0.25Q that has been claimed for the
hole-doped materials (see references \onlinecite{Cai16a} and \onlinecite{Mesaros16a} and below).

Evidence for lattice modulation with {\bf Q} = (0.25, 0.25, 0) was observed by electron diffraction already in 1989,
in both NCCO and Nd$_2$CuO$_{4-x}$F$_x$ near optimal doping \cite{Chen89a}. Superstructure with the same periodicity was also observed
in NCCO by transmission electron microscopy but was ascribed at the time to oxygen vacancies \cite{Aken91a}.
\vskip 0.5pc
\noindent \underbar{(x) Low RE solubility limit.} A remarkable difference between the electron and hole-doped compounds that has {\it not} attracted
the attention it deserves is the low solubility limit of rare earths in the former. While in La$_{2-x}$Sr$_x$CuO$_4$ (LSCO) $x$ can reach as high a
value as 1 in the overdoped region, the upper limit to $x$ in NCCO is about 0.2.
We will argue below that together with all other peculiarities this is also a signature of
valence transition. We also predict similar low electron-dopant solubility in the nominally Ir$^{4+}$ compounds in section~\ref{iridates}.
\vskip 0.5pc
\noindent \underbar{(xi) Zn-substitution effects.} The rapid loss of SC upon Zn-substitution of the Cu ions in the electron-doped superconductors \cite{Barlingay90a}
is arguably one of the most perplexing features of the electron-doped superconductors. Theoretical works on Zn-substitution effects have focused entirely on
hole-doped materials, even as SC vanishes at the same Zn concentration in both
electron and hole-doped systems. In the theoretical literature it is assumed that Zn$^{2+}$ with closed-shell 3d$^{10}$ configuration has the effect of
destroying spin-mediated pairing in the hole-doped materials. This explanation for the destruction of SC cannot
be true for the electron-doped materials, where doping necessarily creates Cu$^{1+}$ with the same 3d$^{10}$ configuration as Zn$^{2+}$.
We discuss the Zn-substitution effect in greater detail in the next subsection.

Observations (iii) - (vii), taken together, point to the same conclusion, viz., carrier generation in the electron-doped
cuprates occurs by a mechanism that is different from simply doping an antiferromagnetic semiconductor.
Observations (vii) - (ix) strongly suggest a massive change in the
electronic structure and orbital occupancy that occurs upon removal of apical oxygens, with Ce-doping acting in a synergistic manner,
that is
not captured in the traditional picture of doping the AFM semiconductor. Observations (viii), (ix) and (xi) indicate that the
electronic structure at
the superconducting composition is likely the same, or at least similar, in the superconducting electron- and
hole-doped superconductors.

\subsection{Experimental puzzles: hole-doped materials}

The experimental and theoretical literature on the hole-doped cuprates are formidably large.
There is a growing consensus that the entry into the PG region at temperature T$^*$ is a true
phase transition and not a crossover \cite{He11a,Shekhter13a,Keimer15a,Sato17a}. The origin of this phase transition is not understood. It is generally believed
that a variety of different broken symmetries, whose natures
are not understood either within existing theories, compete or coexist within the
PG. There is no consensus on whether any of these broken symmetries are the actual drivers of the PG phase transition.
We discuss below what we believe to be the most critical issues.
\vskip 0.5pc
\noindent \underbar{(i) NMR, NQR and Nernst measurements, the case for}
\noindent \underbar{and against fluctuating SC.} Sharp decrease in $^{63,65}$Cu nuclear spin-lattice relaxation \cite{Warren89a} and of static
magnetic susceptibility of the CuO$_2$ plane \cite{Johnston89a,Alloul89a} at T$^*$ gave the first signature of the PG.
One interesting and as yet unexplained phenomenon is the wipeout of Cu-NQR intensity in La-based compounds
upon stripe formation \cite{Hunt99a,Singer99a}.
Fluctuating SC with preformed
spin singlet pairs, as may occur within the resonating
RVB theory \cite{Anderson87a} has been suggested as the possible
origin of reduction of spin susceptibility \cite{Kanigel06a,Chatterjee11a,Dubroka11a,Mishra14a}.
Support for this viewpoint comes from the observation of large positive Nernst signals within the PG region well above T$_c$ but below a temperature
T$_{onset}$ in
underdoped La$_{2-x}$Sr$_x$CuO$_4$ (LSCO), Bi$_2$Sr$_2$CaCu$_2$O$_{8+\delta}$ (Bi2212), Bi$_2$Sr$_{2-x}$La$_x$CuO$_6$ (Bi2201), Bi$_2$Sr$_2$Ca$_2$Cu$_3$O$_{10+\delta}$
(Bi2223) and YBa$_2$Cu$_3$O$_y$ (YBCO) \cite{Wang06a}. Torque magnetometry studies of the same compounds have shown persistence of the diamagnetism
and by implication of local superconducting order up to T$_{onset}$ \cite{Li10b}.

Subsequent experimental work on Bi$_2$Sr$_{2-x}$RE$_x$CuO$_y$ has shown that T$_{onset}$ is significantly smaller \cite{Okada10a} than T$^*$.
Cyr-Choini\'ere {\it et al.} have given an interpretation of the Nernst measurements that is very different
from that in the earliest work \cite{Wang06a,Li10b}, based on experiments
on YBCO, La$_{1.8-x}$Eu$_{0.2}$Sr$_x$CuO$_4$ and La$_{1.6-x}$Nd$_{0.4}$Sr$_x$CuO$_4$ (the experimental observations are the same as before, only the
interpretations are different). The latter authors claim that there are two components to the enhanced
Nernst signal, a magnetic field-dependent quasiparticle contribution due to the reduction in carrier density that occurs at T$^*$, and a second field-independent
contribution at T$_{onset}$ due to pairing \cite{Cyr-Choiniere18a}. Importantly, Cyr-Choini\'ere {\it et al.} argue for a T$_{onset}$ that tracks superconducting
T$_c$ and is much lower than that claimed previously \cite{Li10b}.

T$_{onset}$ significantly smaller than T$^*$ is a signature that pairing is not the
origin of the PG phase transition.
The implicit assumption behind theories suggesting preformed pairs as the origin of PG is that doped cuprates can be described within
single electronic component theory, as in the Zhang-Rice model \cite{Zhang88a}. This assumption has been questioned in recent years
from measurements and analyses of $^{63}$Cu and $^{17}$O NMR shift data \cite{Haase08a,Haase09a,Haase12a} (see also reference \onlinecite{Suter00a}). 
The authors propose a two component model,
one of which is associated with PG behavior, the other with SC. Barzykin and Pines have discussed
a phenomenological two coupled-components model with a spin liquid and a non-Landau Fermi liquid component \cite{Barzykin09a},
the former arising from the Cu $d$-electrons, and the latter from the O $p$-electrons and $d-p$ coupling, respectively.
Whether the two components to the enhanced Nernst signal \cite{Cyr-Choiniere18a} are related to the two components model suggested from
NMR \cite{Haase08a,Haase09a,Haase12a,Barzykin09a} is an intriguiging question. The valence transition model proposed in section ~\ref{model}
presents an integrated microscopic viewpoint of how two distinct components to Nernst and NMR signals emerge.
\vskip 0.5pc
\noindent \underbar{(ii) Spectroscopic signature of anisotropic gap: two}

\noindent \underbar{gaps versus one gap.}
ARPES has been widely used to investigate the energy gap structure of hole-doped cuprates, both in the superconducting phase and in the PG phase.
The overall experimental observations by different groups \cite{Chatterjee10a,Hashimoto14a} are very similar, although controversy persists over the
interpretations of the experiments. The bulk of the experimental works are on Bi2201 and Bi2212. The spectral energy gap in all cases
is dependent on doping, temperature and direction in momentum space. There exist
nodes in the gap with $d_{x^2-y^2}$ structure at the Fermi surface in the superconducting state, and the nodes broaden into so-called Fermi arcs at
finite temperatures. In the near-nodal region (along the diagonal Cu-Cu direction in configuration space)
the gap function is nearly doping-independent and has a simple $d$-wave form.
The gap in the antinodal region (along the Cu-O bond directions in configuration space) (a) continues to exist at temperatures much higher than T$_c$,
(b) is much larger than in the diagonal direction,
and (c) is much larger than that expected from purely $d$-wave behavior, with the deviation larger in the more underdoped systems \cite{Hashimoto14a}.
The antinodal gap
is associated with the PG, and as with the NMR measurements, whether or not this large gap is due to preformed pairs (the one-gap scenario)
or a competing broken symmetry (two gaps) has been a matter of debate \cite{Chatterjee10a,Hashimoto14a}.
The observation of charge- and bond
modulations along the Cu-O directions (see below) would seem to support the second picture. Importantly, the ARPES results support a CO that
extends in both Cu-O directions in a symmetric fashion.
\vskip 0.5pc
\noindent \underbar {(iii) Broken rotational symmetry.} C$_4$ rotational symmetry is broken in underdoped cuprates upon entering the PG phase, and
is replaced with C$_2$ symmetry \cite{Daou10a,Lawler10a,Fujita12a,Kohsaka12a,Achkar16a,Zheng17b,Sato17a}. First observed in the lanthanum family \cite{Tranquada95a},
the phenomenon was originally thought to be associated with a structural low-temperature orthorhombic (LTO) to low-temperature tetragonal (LTT) transition
that confers an apparent 1D character to the system \cite{Axe89a,Hucker10a}. Broken C$_4$ symmetry has also been observed in
Bi2212 \cite{Lawler10a,Daou10a}, Bi$_2$Sr$_2$Dy$_{0.2}$Ca$_{0.8}$Cu$_2$O$_{8+\delta}$ and Ca$_{2-x}$Na$_x$CuO$_2$Cl$_2$ (Na-CCOC) \cite{Kohsaka07a,Kohsaka12a}.
Even in the
(La,M)$_2$CuO$_4$ family it has been found that there is an electronic component of the C$_4$ symmetry breaking that is distinct from the nematicity
induced by structural distortion \cite{Achkar16a}. It is now agreed upon that rotational symmetry breaking is a generic feature of the
underdoped cuprates within the PG phase. In addition, the following observations \cite{Lawler10a,Daou10a,Kohsaka12a} are relevant: (a) nanoscale clusters of localized holes with
C$_2$ symmetry form
immediately upon entering the PG phase in the most highly underdoped cuprates, (b) with increased doping these clusters begin to touch each other
and SC appears at a critical doping level,
(c) the loss of
C$_4$ symmetry is due to {\it electronic inequivalence between the O-ions in the same unit cell} and is
associated with ``weak magnetic states'' on the O-sites \cite{Lawler10a}, and (d) clusters with C$_2$ symmetry are ``aligned'' with the
Cu-O bonds.

Observations (c) and (d) make it unlikely that any simple explanation based on the
idea of domain wall (``stripe'') formation within the antiferromagnetic background will suffice as explanation of the rotational symmetry breaking. A complete
theory should explicitly involve the oxygens, which in turn implicitly supports the two-component scenario suggested by NMR \cite{Haase08a,Haase09a,Haase12a}
and Nernst effect \cite{Cyr-Choiniere18a} measurements. Additional complication arises from the more recent
observations that broken translational symmetry and a consequent charge-ordered phase is also generic to the cuprates in the PG phase (see below)
and that {\it broken translational and rotational symmetry coexist in the hole-doped cuprates} \cite{Fujita14a,Comin15a,Wu15a,Achkar16a}.
It has been shown that in Bi2212 translational and rotational symmetry breakings
vanish at the same critical doping where the full Fermi surface is recovered in ARPES measurement \cite{Fujita14a}.
Finally, the breaking of rotational symmetry is accompanied by a polar Kerr effect \cite{Xia08a,Karapetyan12a,Karapetyan14a,Lubashevsky14a}
that is now believed to be because of 2D chirality and not time reversal symmetry breaking \cite{Karapetyan14a}.
\vskip 0.5pc

\noindent \underbar{(iv) Commensurate doping-independent period 4 CO}. Together with broken rotational symmetry, it is by now widely accepted that CO
is a generic feature of the
underdoped and optimally doped cuprates \cite{Hoffman02a,Howald03a,Momono05a,Hanaguri04a,Shen05a,Liu07a,Hucker11a,Kohsaka12a,LeTacon14a,Wu13a,Wu15a,Tabis14a,Blanco-Canosa13a,Blanco-Canosa14a,Hashimoto14b,Comin14a,Comin15a,SilvaNeto14a,Peng16a,Cai16a,Mesaros16a}. First observed in the La-based compounds it has now been seen in all the
superconducting cuprates (including the electron-doped materials, see previous subsection). Following intense investigations by many experimental groups, a
number of highly specific observations that appear to be true for all the cuprates have emerged. These are listed below.

(a) The charge modulation is overwhelmingly on the layer O-ions \cite{Wu15a,Comin15a}. 
In particular, NMR experiments find only two kinds
of oxygens \cite{Wu15a}, which likely indicates nominal valence states O$^{1-}$ and O$^{2-}$, and not multiple valences.

(b) The charge modulation is accompanied by bond order modulations along the Cu-O bond directions 
(we will argue below that the bond modulations involve the O-Cu-O linkages) and exhibits a ``$d$-wave pattern'' \cite{Liu07a,Comin14a,Comin15a,Hashimoto14b,Hamidian15a}.

(c) The charge modulation and the C$_4$ rotational symmetry breaking with inequivalent intraunit cell (IUC) oxygen ions
appear at the same temperature T$_{CO}$ in underdoped materials \cite{Kohsaka12a,Wu15a,Comin15a} and disappear
at the same critical high dopant concentration \cite{Fujita12a,Fujita14a}. It is therefore believed that the IUC {\bf Q} = 0 symmetry
breaking is a consequence of the {\bf Q} $\neq 0$ CO. T$_{CO}$ is also the same temperature where the polar Kerr
effect appears \cite{Wu15a,Karapetyan14a}. It is likely that all three phenomena,
C$_4$ rotational symmetry breaking, CO and the Kerr effect are intimately coupled.

(d) Not only does the CO does not coexist with AFM \cite{Wu15a}, it (as well as the PG phase itself) is also easily destroyed by
Zn-doping \cite{Blanco-Canosa13a,Hamidian16a}. We will return to this below.
\begin{table*}
\centering
\caption{Charge-order characteristics of hole- and electron-doped cuprates. The astersisks against specific doping
concentrations indicate superconducting compositions. The third column gives the CO symmetries as described by the authors
of the experimental papers.}
 \begin{tabular}{ p{3cm}  p{3cm}  p{3cm}  p{2.5cm}  p{4cm}  }
 \hline
 Material & Doping & CO symmetry & CO periodicity & Experimental technique\\ [0.5ex]
 \hline\hline
Na-CCOC & 0.08,0.10,0.12* & "checkerboard" & 0.25Q & STM  \cite{Hanaguri04a}\\
  & 0.05,0.10,0.12* & "2D" & 0.25Q & ARPES \cite{Shen05a}\\
\addlinespace[0.2cm]
Bi2212 & optimal & "checkerboard" & 0.25Q & STM \cite{Hoffman02a}\\
  & optimal & "checkerboard" & 0.25Q & STM \cite{Howald03a}\\
  & $<$0.1 & "(Q$^*$,0);(0,Q$^*$)" & $\sim$ 0.3Q & STM, RXS \cite{SilvaNeto14a}\\
  & $>$0.1 & "(Q$^*$,0);(0,Q$^*$)"  & $\sim$ 0.25Q & STM, RXS \cite{SilvaNeto14a}\\
  & 0.06,0.08,0.10, & "(Q$^*$,0);(0,Q$^*$)" & 0.25Q & STM \cite{Mesaros16a}\\
  & 0.14*,0.17* & "$d$-density wave form factor" & 0.25Q & STM \cite{Mesaros16a}\\
\addlinespace[0.2cm]
Bi2201 & 0.115,0.130,0.145* &"(Q$^*$,0);(0,Q$^*$)" & 0.243-0.265Q & RXS,STM,ARPES \cite{Comin14a,Comin15a}\\
  & 0.07-0.16* & -- & 0.26-0.23Q & RIXS \cite{Peng16a}\\
  & 0.03,0.07,0.10* & "checkerboard" & 0.25Q & STM \cite{Cai16a}\\
\addlinespace[0.2cm]
 Pb-Bi2212 & optimal* & -- & $\sim$ 0.28Q & RIXS \cite{Hashimoto14a}\\
\addlinespace[0.2cm]
 Hg-1201 & $\sim$ 0.09 & "checkerboard" & $\sim$ 0.27-0.28Q & RXD,RIXS \cite{Tabis14a}\\
\addlinespace[0.2cm]
YBCO:LCMO & $\sim$ 0.1* & "(Q$^*$,0)" & 0.245Q & RXS \cite{He16a} \\
NCCO & 0.14$\pm$0.01* & -- & (0.23$\pm$0.04)Q & {RXS \cite{SilvaNeto15a}, RSXS \cite{Jang17a}}\\
  & 0.15 $\pm$0.01* & -- & (0.24$\pm$0.04)Q\\
\addlinespace[0.2cm]
LCCO & 0.08* & --& $\sim$ 0.22 & RXS \cite{SilvaNeto16a} \\
 \hline
\end{tabular}
\end{table*}
Beyond the above, complete characterization of the CO requires knowledge of its doping dependence, periodicity and symmetry. There is now increasing
evidence that the periodicity is universally  0.25Q. 
This periodicity is seen in all La-based compounds at the lowest temperatures. However, 
the density wave here has most often been described as 1D stripes. Based on the behavior of the other cuprates (see below), we believe that the
true structure of the CO in the La-based materials is obscured by the LTO-to-LTT transition \cite{Axe89a,Hucker10a}. Our discussions of the La-based materials
therefore will be limited. 

In Table 1 below we have listed recent experimental results for
Na-CCOC, Bi2212, Bi$_2$Sr$_{2-x}$La$_x$CuO$_{6+\delta}$
(La-Bi2201), Bi$_{1.5}$Pb$_{0.6}$Sr$_{1.54}$CaCu$_2$O$_{8+\delta}$ (Pb-Bi2212), HgBa$_2$CuO$_{4+\delta}$ (Hg1201),
YBCO thin films grown epitaxially on La$_{0.3}$Ca$_{0.3}$MnO$_3$ (LCMO) \cite{He16a}, and electron-doped NCCO and LCCO.
In each case we have given the
doping range for which the experiments were performed, the periodicities, the symmetries of the CO {\it as described by the authors of the experimental
investigations} and the experimental
techniques that have been used to detect the CO.
As seen from the Table, {\it doping-independent commensurate periodicity} of exactly 4a$_0$ is the most likely outcome. Indeed, this commensurate periodicity, independent of doping,
has been found in Na-CCOC \cite{Hanaguri04a,Shen05a}, La-Bi2201 \cite{Cai16a} and Bi2212 \cite{Hoffman02a,Howald03a,Liu07a,Mesaros16a}. Deviations from
commensurability are weak in
all cases shown in Table 1. It has been argued that weak deviations seen here are due to discommensurations
within a commensurate CO background \cite{Mesaros16a} that render an {\it apparent} incommensurate character to the CO whose
fundamental wavevector is however 0.25Q.
Although several earlier studies on YBCO \cite{Ghiringelli12a,Blackburn13b,Blanco-Canosa14a,Hucker14a} found CO wavevector closer to 0.3Q, a NMR study has indicated
commensurate CO \cite{Wu15a} with periodicity 4a$_0$; the latter periodicity has also been observed in the YBCO:LCMO heterostructure \cite{He16a}.
Remarkably, {\it the CO periodicities in the optimally electron-doped NCCO and LCCO are virtually the same as in the hole-doped systems}
\cite{SilvaNeto15a,SilvaNeto16a,Jang17a}. It has been argued that the CO's in the hole and electron-doped cuprates are different in
character \cite{SilvaNeto16a,Jang17a}. The close matching of the CO wavevectors in this case will have to be a coincidence. Within the valence
transition model the same CO wavevector is predicted (see below).

While a consensus is thus emerging on the periodicity of the CO, the discussion of the symmetry has been somewhat confusing. The CO has been
described both as 2D ("4a$_0$ $\times$ 4a$_0$" or "checkerboard") \cite{Hanaguri04a,Shen05a,Cai16a} as well as "(0.25Q,0);(0,0.25Q)"
\cite{Hanaguri04a,Comin15a,Mesaros16a}. It is not entirely clear whether the latter classification has been meant to imply 2D CO or 1D stripes, as the
corresponding experiments have often found evidence of modulations along both the Cu-O directions (but not the diagonal Cu-Cu direction).
The most likely explanation is that experiments overwhelmingly detect bond order modulations rather than charge modulations
(although in a non-$\frac{1}{2}$-filled band they accompany each other \cite{Clay03a})
and the bond order modulations, which occur along both the Cu-O axes but not the diagonal Cu-Cu direction, appear as
interpenetrating stripe-like structures \cite{Howald03a}.
This would explain the symmetry between the two axes implied in ARPES measurements. We will show that precisely such a 2D CO with
period 4 bond modulations along both Cu-O directions is expected within the present theory.

Doping-independent commensurate periodicity precludes the possibility that the CO is a consequence of nesting, and suggests that
the mechanism behind the CO formation should be found from configuration space arguments \cite{Cai16a,Mesaros16a}.
One additional important point is that CO with the same
periodicity in superconducting samples \cite{Cai16a} may suggest possible coexistence of CO and SC, which may indicate the CO is a {\it density wave of
Cooper pairs} \cite{Anderson04b,Franz04a,Tesanovic04a,Chen04a,Vojta08a,Hamidian16a,Cai16a,Mesaros16a}, a possibility that we will return to later.
\vskip 0.5pc
\noindent \underbar {(v) Strong electron-phonon coupling, giant phonon}
\noindent \underbar { anomaly.} There is now strong experimental evidence for
giant softening of the Cu-O bond stretching phonon frequency in the underdoped cuprates \cite{Reznik06a,Reznik10a,LeTacon14a,Park14a,Peng16a}.
Periodicity $\sim$ 0.25Q is again observed most commonly.
Reznik {\it et al.} have repeatedly emphasized the role of the so-called
half-breathing mode \cite{Reznik06a,Reznik10a,Park14a}.
The apparent similarity as well as differencees with the traditional Kohn anomaly observed in 1D charge-density wave systems
has been noted \cite{Reznik10a}.
On the one hand it is clear that the phonon anomaly is related to the CO formation
discussed above. On the other hand, it is unlikely that the relatively weak electron-phonon coupling is the main driver of the CO and SC.
The most likely cause of the phonon anomaly is then co-operative coupling, with electron-electron interactions driving the CO instability,
and the phonon softening occurring as a consequence of the same \cite{Clay03a}.
\vskip 0.5pc
\noindent \underbar{(vi) Zn-substitution effects on SC, CO and PG.} The effect of substituting Zn for Cu is dramatically
deleterious to SC and is also generic to all the hole-doped materials
\cite{Mahajan94a,Mizuhashi95a,Fukuzumi96a,Bernhard96a,Nachumi96a,Julien00a,Pan00a,Itoh03a,Adachi04a,Pelc15a}. In spite of the
nonmagnetic character of the Zn$^{2+}$ cation there is drastic reduction of T$_c$ upon doping with few percent of
Zn. The overall experimental results, obtained by using a variety of experimental techniques, can be summarized as
follows.

(a) The reduction in T$_c$ is due to severe decrease in superfluid density around each impurity ion \cite{Nachumi96a}. The reduction
of superfluid density in YBCO$_{6.6}$ is 70\% for Zn doping concentration of 2\%.

(b) There occur {\it insulating} islands
with spontaneous phase separation between superconducting and nonsuperconducting regions in the material, with
the regions with charge localization characterized by simultaneous
staggered magnetization about the impurity centers and and enhancement of the antiferromagnetic correlations
(this is sometimes referred to as the Swiss cheese model of exclusion of superfluid density).

(c) Zn-substitution is equally deleterious to the CO within the PG state,
leading again to enhancement of incommensurate spin correlations \cite{Blanco-Canosa13a,Hamidian16a}.
The spin gap that is seen in the PG region of the underdoped materials
either vanishes or is filled in.
The bulk spin susceptibility shows Curie-like behavior, as if the moments around the dopant centers are noninteracting.

Theoretical efforts to explain the deleterious effect on SC have focused on the spinless character of the Zn$^{2+}$ ion which in principle
acts as a vacancy within the RVB model or spin fluctuation theories, thus having a pair-breaking effect. While enhancement of
local spin moments around the vacancy center is to be expected, it is more difficult to understand the charge localization and phase separation
if Zn$^{2+}$ ions behave simply as spinless vacancies.
Enhanced antiferromagnetic correlations is also difficult to understand, given the Zn-doping of the parent semiconductor is deterimental to AFM \cite{Cheong91a}.
{\it The detrimental effect of Zn-doping on the PG state and CO \cite{Blanco-Canosa13a,Hamidian16a} is particularly
perplexing, because the AFM and CO states are both charge-localized.} This last observation indicates that Zn-doping simultaneously
destroys spin pairing and prefers one kind of charge-localized state over another!
Finally, it is not at all possible to
understand the disappearance of SC by Zn doping
in the electron-doped materials \cite{Barlingay90a}, since the electron configurations of Zn$^{2+}$ and that of Cu$^{1+}$ are identical. The simultaneous
charge localization and enhancement of spin moment on Cu-sites clearly indicates a coupled charge-spin mechanism as opposed to purely spin-only mechanism.

\subsection{Mechanism of SC.}

Approximate theories of the weakly doped Mott-Hubbard semiconductor, with carrier concentration $\rho$ in the range of 0.8 - 0.9 electrons per site, often find the
system to be superconductng, either within
the single-band Hubbard model, or within the three-band model. DMFT calculations, in particular, find SC
within the weakly-doped single-band Hubbard model \cite{Maier04a,Khatami10a,Capone06a,Gull12a,Sordi12a,Fratino16a,Tocchio16a}. Yet quantum Monte
Carlo (QMC) or path integral renormalization group (PIRG) calculations that have searched for long-range superconducting
pair-pair correlations have consistently found suppression of superconducting pair-pair correlations
by repulsive Hubbard U in the same carrier concentration range \cite{White89a,Zhang97a,Chang08a,Chang10a,Aimi07a,Misawa14a}, casting doubt on the
DMFT results. QMC calculations are mostly for relatively small Hubbard $U$ ($U \leq 4|t|$, where $t$ is one-electron hopping).
Recently several authors using the variational Monte Carlo
\cite{Yokoyama13a,Yanagisawa16a,Tocchio16a} and dynamic cluster approximation (DCA) \cite{Gull13a}
have suggested that SC occurs within the $\rho\sim 0.8$ Hubbard model
only when $U>U_c$, with $U_c \sim $ 4--6 $|t|$.
However, a recent calculation of $d_{x^2-y^2}$ pairing correlations in $\rho=0.875$ for Hubbard cylinders of width 4 and 6 sites, using a hybrid
real-momentum space formulation of the density matrix renormalization group (DMRG) approach has found that pairing
correlations decay exponentially with distance \cite{Ehlers17a}, even for $U$ as large as 8$|t|$.

Mean field and DMFT approaches that find SC within the weakly doped Hubbard model on a square lattice
also find SC in the
anisotropic triangular lattice for the exactly $\frac{1}{2}$-filled Hubbard band \cite{Kyung06a,Yokoyama06a,Sahebsara06a,Sentenf11a,Hebert15a},
a theoretical model often assumed for the organic charge-transfer solids (BEDT-TTF)$_2$X which exhibit pressure-induced AFM-to-SC transition at constant filling.
Once again, suppression of superconducting pair-pair correlations by Hubbard $U$ is found from
exact diagonalization \cite{Clay08a,Gomes13a} and PIRG studies \cite{Dayal12a}. The likely reason for this discrepancy between DMFT and DCA on the one hand and
QMC and DMRG on the other
is that pair correlations are indeed enhanced by Hubbard $U$ for overlapping pairs at short interpair distances, where antiferromagnetic correlations
contribute to the enhancement, but are suppressed at larger
interpair distances \cite{Dayal12a,Moreo92a}. This effect is perhaps not captured in small cluster DMFT calculations, especially in the absence of
calculations that do not separate out short versus long-range pair-pair correlations.

In addition to the above, the weakly doped Mott-Hubbard semiconductor model of superconducting cuprates does not capture the normal state behaviors
described in sections IIA and IIB. There is so far no consensus on whether or not there occurs a phase transition to the
PG phase within the doped Hubbard model.
Recent theoretical calculations of stripe order within the Hubbard model for $\rho=0.875$ using multiple different techniques {\it do not} find
charge modulations with periodicity 4$a_0$,
as would be required from Table I \cite{Zheng17a}. The authors conclude that the 2D Hubbard model may not be appropriate for cuprates.

\subsection{Summary}

Any comprehensive theory of cuprates must explain the simultaneously remarkable similarities and differences between the electron- and hole-doped
materials. While the similarities arise from the underlying CuO$_2$ layers that are common to both, the differences must arise from the differences
in the crystal structures. The latter already suggests that Madelung energy considerations \cite{Mazumdar89a,Mazumdar89b} are important.
The discovery that O-ions play very significant roles in the breaking of both translational and rotational symmetries presents us with a
very difficult conundrum. On the one hand these experiments indicate that any single-band Cu-based model for hole-doped materials is
insufficient. On the other, only single-band models can simultaneously explain CO and SC in both hole- and electron-doped cuprates!
In the remaining of the paper, we discuss the valence transition model, in which O-ions and
the Madelung energy play dominant roles, and the application of the model to cuprates.

\section{The valence transition model}
\label{model}

We are interested in the true ionicities in the CuO$_2$ layer, as a function
of doping. The rare earth, bismuth and mercury ion valencies in the cuprates are taken to be independent of doping, which is a safe assumption. It is agreed upon at
the outset that even if the true Cu ion charges are not exactly integral, the physical behaviors can mimic those expected from integer charges, provided the
true charges are close enough.
Note that this assumption is built into the superexchange model of AFM in the undoped state, where the true ionicity of the Cu ions is likely
less than +2, yet it is sufficiently close that the O-ions (also with true charge less than the formal charge -2) behave as closed-shell O$^{2-}$,
giving validity to the effective single $\frac{1}{2}$-filled Cu-band Mott-Hubbard description that we are familiar with. We will argue that there exists another
distinct and proximate state in which the true Cu-ion charge is siginificantly less than +2 and
close enough to +1 that it behaves as closed-shell Cu$^{1+}$.
Should transition to this state with ``negative charge-transfer gap'' occur the charge-carriers would occupy an effective $\frac{1}{4}$-filled
band of O$^{1-}$ holes ($\frac{3}{4}$-filled electron band), with the closed-shell Cu$^{1+}$ ions now as inactive as the closed-shell O$^{2-}$ ions are in the AFM semiconductor. Below we discuss
the mechanism of such a first order phase transition. While our focus is on the cuprates our discussion below is in the context of transition metal oxides
in general, both for clarifying why Cu is special in the 3$d$ series, and for later application to SrIr$_2$O$_4$.

We begin with the ZSA scheme for classification of transition metal oxides \cite{Zaanen85a} and consider the competition between electron configurations
$d^np^6$ and $d^{n+1}p^5$, versus the Mott-Hubbard energy gap $U_d=d^n + d^n \to d^{n+1} + d^{n-1}$. We define the charge-transfer
gap in the usual manner, $\Delta$ = E($d^{n+1}p^5$) -- E($d^np^6$), where E($\cdots$) is the total ground state energy of the state.
Then for $|\Delta| > U_d$ the system is a Mott-Hubbard
insulator, for $|\Delta| < U_d$ the system is a charge-transfer insulator \cite{Zaanen85a}. Note that
the system remains a charge-transfer insulator even if the sign of $\Delta$ is negative
\cite{Mizokawa91a,Khomskii01a,Green16a,Plumb16a}.
We will not only be interested in when this is most likely, but whether there can occur a real transition from positive to negative $\Delta$.
\subsection{The role of the second ionization energy}

We continue this discussion from a strong correlations perspective (as opposed to one based on band theoretical considerations) 
that recognizes at the outset the atomistic and many-body contributions to 
E($d^np^6$) and E($d^{n+1}p^5$), 
as is done in the context of the neutral-to-ionic transition in charge-transfer solids
\cite{Torrance81a,Torrance81b,Tokura89a,Kosihara90a,Horiuchi03a,Mazumdar78a,Pati01a,Nagaosa86a}. We write I$_n$ as the $n$th ionization energy of the 
transition metal M [M$^{(n-1)+} \to$ M$^{n+}$ + e],
A$_2$ as the second electron affinity of oxygen [the energy needed to add the second electron to neutral oxygen, O$^{1-} + e \to$ O$^{2-}$], and
E$_{M,n}$ as the Madelung energy  stabilization of the solid with the cation in the charged state M$^{n+}$. 
In the limit of small electron hopping $t_{pd}$ between the cation and oxygen, the inequality
\begin{equation}
I_n + A_2 + \Delta E_{M,n} + \Delta(W) \gtrless 0
\label{ionicity}
\end{equation}
determines the actual ionicity of the transition metal oxide layer,
where $\Delta E_{M,n}$ = E$_{M,n}-$E$_{M,n-1}$, and $\Delta(W)=W_n-W_{n-1}$, with $W_n$ ($W_{n-1}$) the gain in electronic energy due to 
electron delocalization
with all cations as M$^{n+}$ (M$^{(n-1)+}$). 
Smaller left hand side in Eq.~\ref{ionicity} implies positive charge-transfer gap with the metal ion as M$^{n+}$; larger left hand side means 
negative charge-transfer gap and the metal ion is in the state M$^{(n-1)+}$. 
Note that the occurrence of two distinct states with nearly integer valences \cite{Torrance81a,Torrance81b,Tokura89a,Kosihara90a,Horiuchi03a,Mazumdar78a,Pati01a,Nagaosa86a}, 
as opposed to mixed valence or covalency, 
is a consequence of strong correlations relative to $t_{pd}$. In the specific
case of the competition between Cu$^{2+}$ versus Cu$^{1+}$ in the oxide the tendency to covalency is particularly weak, since
{\it the covalent bond would result from sharing a single hole, as opposed to a Lewis pair, between the two constituents.} 
Only one of the two constituent ions in either of the
states Cu$^{2+}$O$^{2-}$ and Cu$^{1+}$O$^{1-}$ is closed-shell; for strong covalency it is required that both constituents
are closed shell in at least one of the configurations.

For the layered cuprates, the following are relevant.  
(i) A$_2$ is positive (it costs energy
to add the second electron to O$^{1-}$) because of the Coulomb repulsion between like charges in O$^{2-}$, even as the first electron affinity of oxygen is negative.
(ii) E$_{M,n}$ and E$_{M,n-1}$ are three-dimensional (3D) even if electron or hole-motion is in the 2D CuO$_2$ layer. {\it This observation has an important   
implication of its own, viz.,  in the ``metallic'' or superconducting states the 3D Madelung energy should not pin the charge in the CuO$_2$ layers to specific sites outside the layers.
Conversely, the true charges of the ions in the layers must necessarily be impervious to the pinning effect due to the 3D Madelung energy.}
(iii) While $\Delta E_{M,n}$ is necessarily negative, favoring higher positive charge on M (and all O-ions as O$^{2-}$),
with twice as many O as Cu ions in the CuO$_2$ layer, we anticipate $W_{n-1}$ to be more negative
than W$_{n}$, simply because of a larger number
of charge carriers when 50\% of the oxygens are O$^{1-}$. High charge is therefore favored only by  $\Delta E_{M,n}$. The overall competition between M$^{n+}$ and
M$^{(n-1)+}$ is then determined almost entirely by the relative magnitudes 
of the two largest quantities, $I_n$ and
$\Delta E_{M,n}$, with the other two terms both favoring {\it lower} charge. 

\begin{figure}
\includegraphics[width=3.0in]{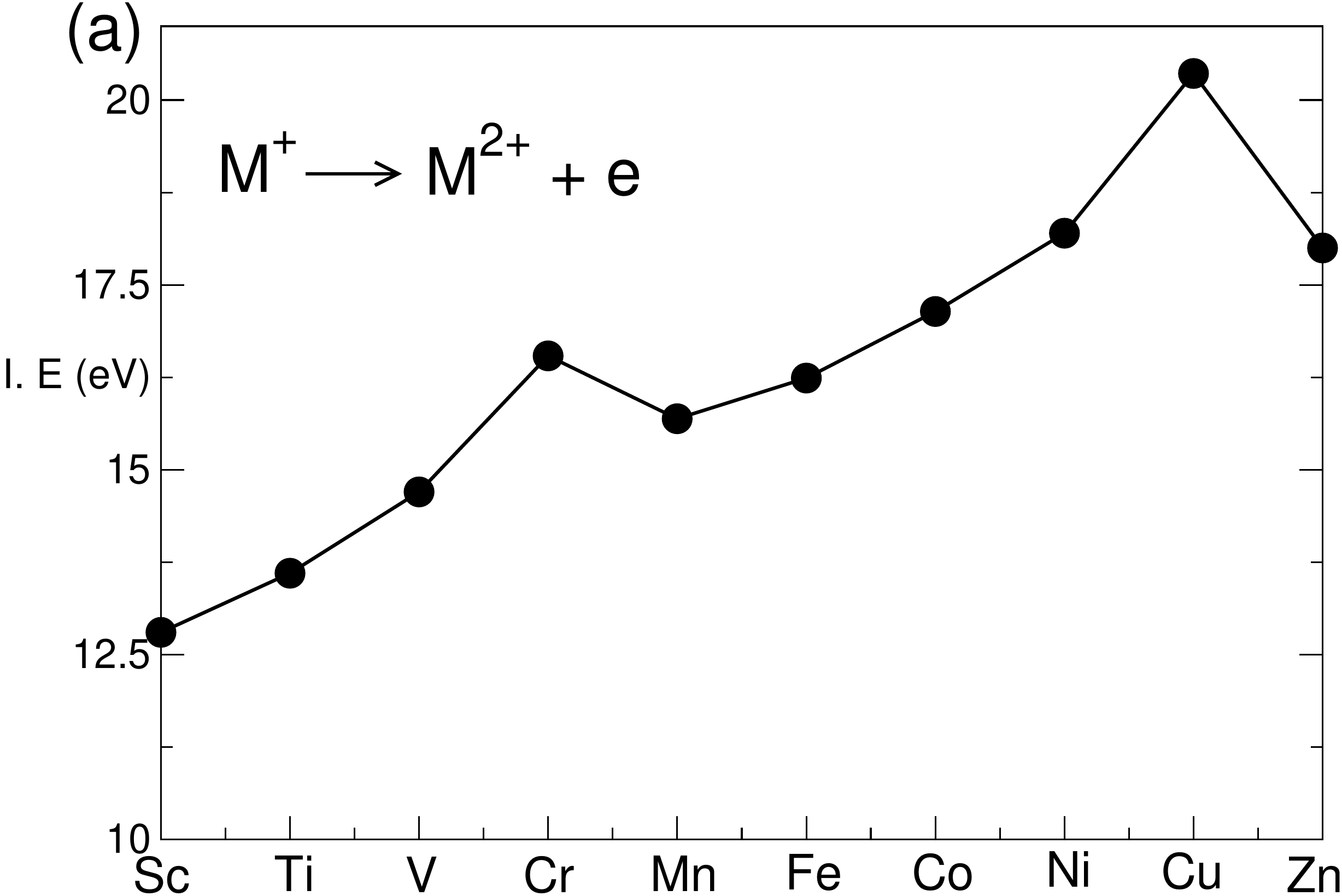}
\caption{Second ionization energies of 3$d$ transition metals.}
\label{IE}
\end{figure}

In Fig.~\ref{IE} we have plotted the second ionization energy I$_2$ (M$^+$ $\to$ M$^{2+}$ + e) of the first row transition metals. {\it I$_2$ 
of Cu is the largest in the series,} larger than those of Ni and Zn by $\ge 2$ eV, because of the closed shell (3d$^{10}$) nature of Cu$^{1+}$ (the smaller
peak at Cr is due to the $\frac{1}{2}$-filled $d^5$-occupancy of Cr$^{1+}$ with strong Hund's rule coupling). The large decrease of I$_2$ 
of Zn is similarly
due to the closed-shell (and hence highly stable) nature of Zn$^{2+}$. {\it Thus is it only the gain in Madelung energy that gives the Cu$^{2+}$(O$^{2-}$)$_2$
electronic configuration of the parent cuprate semiconductors.} 
Based on the above discussions and Fig.~\ref{IE} we posit that the 
undoped cuprates are very close to the boundary between the two phases with charges Cu$^{2+}$ and Cu$^{1+}$. Doping (or O-deficiency in the
case of electron-doped materials, see (iv) and (v) in section IIA) can therefore lead to discrete jump in ionic charge from Cu$^{2+}$ to Cu$^{1+}$ 
(with nearly half the oxygens in the state O$^{1-}$)
due to reduction in the magnitude of $\Delta E_{M,n}$ and contribution from $\Delta(W)$.
{\it It is reemphasized that similar preponderance of O$^{1-}$ has been recently confirmed in (RE)NiO$_3$ and BaBiO$_3$.}
The reasons behind the lower ionicities in these are the same, very high third and fourth ionization energies of Ni and Bi, respectively. Bi$^{3+}$,
in particular, is closed shell, exactly as Cu$^{1+}$, conferring it extra stability.
 
\subsection{Charge introduction in the CuO$_2$ layers, T versus T$^{\prime}$ structures.}
\vskip 0.5pc
\begin{figure}
\includegraphics[width=3.5in]{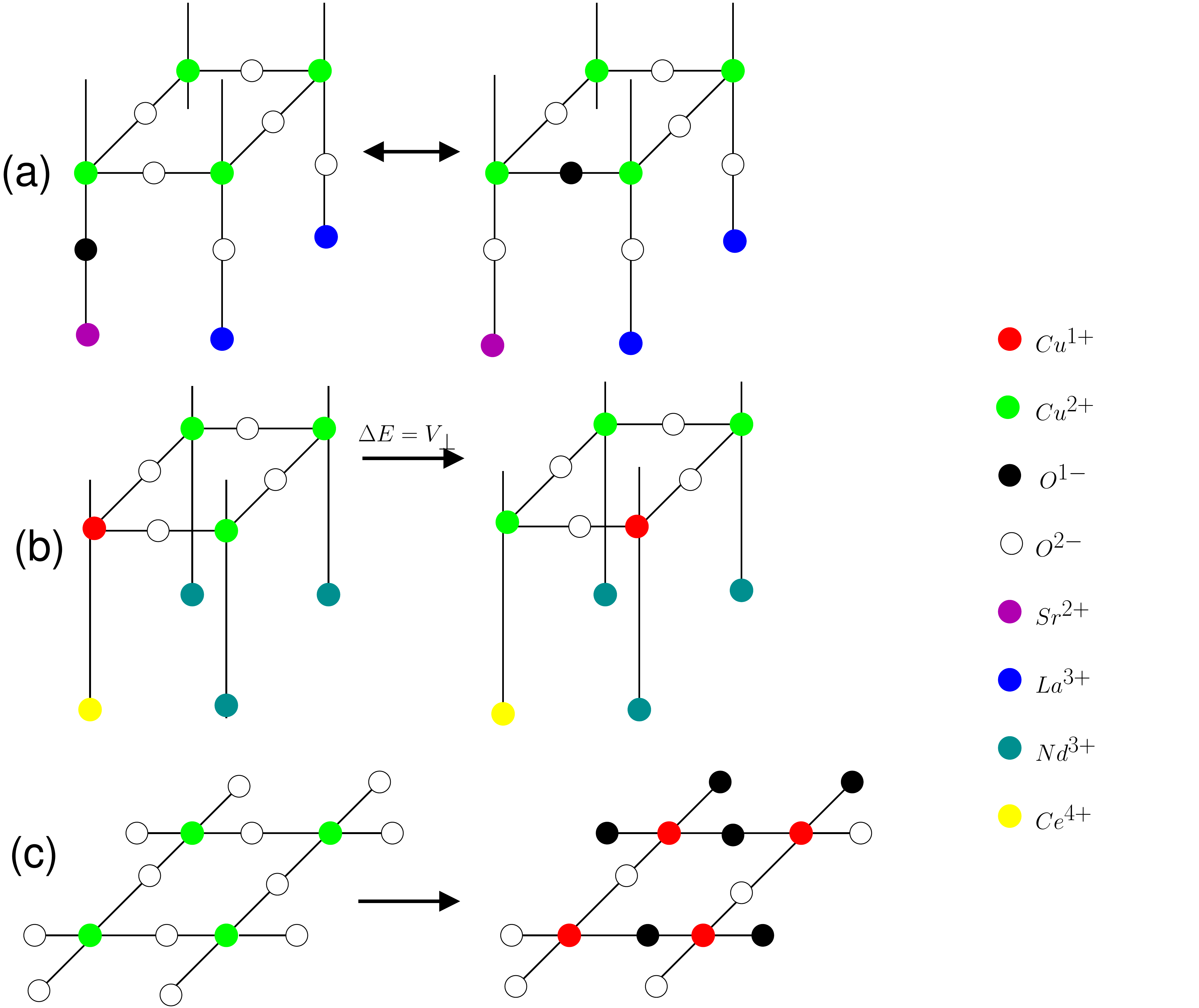}
\caption{(a) Schematic of hole doping by chemical susbstitution in LSCO. Substitition of La$^{3+}$ with Sr$^{2+}$ in the T-structure can 
generate an O$^{1-}$ in the apical position or in the CuO$_2$ layer. The Madelung energies of the two configurations with O$^{1-}$ in the different positions are nearly the same. 
The energy barrier to the delocalization of the hole in the layer is the weak long range component of the 3D Coulomb interaction, giving the 2D 
polaronic ``bad metal''. 
(b) Schematic picture of electron doping in the T$^{\prime}$-structure of NCCO.
The electron on the Cu$^{1+}$ is pinned to Ce$^{4+}$, as electron motion would yield nearest neighbor Ce$^{4+}$--Cu$^{2+}$ with 
strong nearest neighbor Coulomb repulsion. The charge localization leaves the system in the semiconducting antiferromagnetic state with reduced spin moment.
(c) Valence transition driven by dopant-induced reduction in the Madelung energy, in both hole- and electron-doped materials. The 3D
Madelung energy barrier to 2D 
charge motion is now absent in both classes of materials.}   
\label{ehdoped}
\end{figure}

\noindent We show in this subsection how the difference between the weakly doped hole- and electron-doped cuprates, the ``bad metal'' phase in the
former that occurs above the PG phase transition and the stable AFM that persists upto optimal doping in the conventionally prepared electron-doped
materials both fit in with this theoretical picture.
In Figs.~\ref{ehdoped}(a) and (b) we have shown an atomistic picture of ``doping'' -- introduction 
of charge in the CuO$_2$ layer by chemical substitution away from the layers --  prior to the
valence transition, in the T versus T$^{\prime}$ structures. As shown schematically in Fig.~\ref{ehdoped}(a) 
there is little to no difference in the Madelung energy stabilization between configurations with the doped hole on the apical versus layer O, 
as the dominant contributions to the Madelung energies of configurations Sr$^{2+}$--O$^{1-}$--Cu$^{2+}$ and Cu$^{2+}$--O$^{1-}$--Cu$^{2+}$ are nearly equal 
(the closest La$^{3+}$ ions are nearly equally far in both cases) \cite{Torrance89a}. The gain in delocalization energy due to band motion places the hole preferably in the
2D CuO$_2$ layer, in agreement with observations. The Madelung energy barrier to the charge moving to the next more 
distant O within the layer is the {\it difference} between the third- and fifth-neighbor Coulomb interaction (see Fig.), much but not all of which is `screened out''.
The resultant state is the ``bad metal'' in hole-doped cuprates at T $>$ T$^*$. The above scenario applies to all families with apical O.

The consequence of chemical substitution in the conventional T$^{\prime}$ structure is drastically different, as is shown schematically in Fig.~\ref{ehdoped}(b).
Replacement of a Nd$^{3+}$ with Ce$^{4+}$ adds an electron on the nearest Cu$^{2+}$, converting it to Cu$^{1+}$. {\it The doped electron is pinned to this 
particular Cu-ion},
since a nearest neighbor hopping of the electron to any of the neighboring Cu-ions would generate nearest neighbor Ce$^{4+}$- Cu$^{2+}$ with the very large 
Madelung repulsion. The persistent commensurate AFM in the conventional electron-doped cuprates is due to this 
Madelung energy driven charge pinning in the less than optimally doped $T^{\prime}$ structure, 
and not because of any difference between the Hubbard $U$ on Cu-ions in electron versus hole-doped compounds. The N\'eel temperature T$_N$ is reduced due to localized defects \cite{Cheong91a}.
The semiconducting behaviors of conventional NCCO and PCCO for $x \leq 0.13$ are then to be expected. 
As already mentioned, metallicity and SC require the 3D Coulomb interaction between the dopant ion outside the layer and the ions in the layer to become ``irrelevant''. 
This can happen only following the valence transition (see below).

\subsection{Valence transition and the effective Hamiltonian}

\noindent With increasing  number of charges introduced in the layers, there occurs reduction in the magnitude of $\Delta E(M,n)$ and enhancement of $\Delta(W)$. 
Recall that we have argued that given the large I$_2$ of Cu, only the preponderance of O$^{2-}$ gives large enough $|\Delta E(M,n)|$ for the Cu-ions to be in the +2 state. The
valence transition that occurs in the CuO$_2$ layers in the PG phase of hole-doped compounds and in the optimally electron-doped conventional $T^\prime$ 
compounds layers is the same and is shown in Fig~\ref{ehdoped}(c). The true Cu-ion charges are + (1.0 + $\epsilon$), where $\epsilon(x)$
remains small throughout the PG phase, such that the Cu-ions behave as nonmagnetic closed-shell Cu$^{1+}$. The physical consequence of 
nonzero $\epsilon(x)$ can be ignored, the logic being the same as for ignoring the weak deviation from the exact charge of +2 in the antiferromagnet. This first order
transition is the PG phase transition in the hole-doped materials. We postpone the discussion of the dependence of T$^*$ on doping, until after we have discussed the
commensurate period 4 CO in the next section. Given that half the O-ions are now hole carriers, $\Delta(W)$ is now much larger, also favoring the
monovalent Cu-ion conguration.
Not only the stable AFM is lost in the electron-doped materials, but also i{\it the 3D Coulomb barrier
to layer charge motion (O$^{1-}$-Cu$^{1+}$-O$^{2-} \to$ O$^{2-}$Cu$^{1+}$O$^{1-}$) is now nonexistent:} with all
the Cu-ions monovalent and the oxygens as O$^{1.5-}$ the different 3D configurations have nearly the same Madelung energies. 
Within the valence transition mechanism the semiconducting
AFM phase behaves as an effective  $\frac{1}{2}$-filled Hubbard band with the spins on the Cu sites; the O$^{2-}$-sites are
inactive because of their closed-shell nature. The new proposition is that in the
PG and superconducting states the active sites consist of
nearly $\rho=0.5$ $p$-band with all  
charge carriers on the O$^{1.5-}$ sites; now the Cu$^{1+}$ sites are inactive because of their closed-shell configuration. The effective
Hamiltonian of the optimally doped electron-doped cuprates, and within the pseudogapped region of the hole-doped
cuprates is the same, viz.,
\begin{eqnarray}
 H = -\sum_{[ij],\sigma}t_p (p^\dagger_{i,\sigma}p_{j,\sigma}+H.c.) + U_p\sum_i n_{pi,\uparrow}n_{pi,\downarrow} \\
\nonumber
+ \frac{1}{2}\sum_{\langle i j\rangle} V_p^{NN} n_{pi} n_{pj} + \frac{1}{2}\sum_{(i j)} V_p^{NNN} n_{pi} n_{pj}\\
\nonumber
\label{hamuv}
\end{eqnarray}
where the sums are over the O-ions in the CuO$_2$ layer, $p^\dagger_{i,\sigma}$ creates a charge carrier on an O$^{2-}$ ion to create O$^{1-}$,
$n_{pi,\sigma}=p^\dagger_{i,\sigma}p_{i,\sigma}$ and 
$n_p=\sum_{\sigma=\uparrow,\downarrow}n_{pi,\sigma}$. Note that this convention (O$^{2-}$ as the vacuum) gives a $\frac{1}{4}$-filled (as opposed to $\frac{3}{4}$-filled)
description. Here ${[ij]}$ implies O $p$-orbitals linked through the same Cu$^{1+}$.
The O-sublattice is a strongly frustrated checkerboard lattice with $t_p$ for the
O-Cu-O carrier hoppings the same, irrespective of whether the O-Cu-O bond angle is 90$^o$ or 180$^o$.
Here $t_p = t_{dp}^2/\Delta E$, $\Delta E$ = E(Cu$^{1+}$O$^{1-}$)$-$E(Cu$^{2+}$O$^{2-}$),
where E($\cdots$) is the energy of the corresponding configuration embedded in the background with negative charge-transfer gap.
Direct O-O hopping can also be included for O-O hoppings with 90$^o$ Cu-O-Cu bond angle, but this is not essential 
for what follows. $V_p^{NN}$ and $V_p^{NNN}$ are the effective Coulomb repulsions between charge carriers on nearest neighbor O-ions 
(linked by 90$^o$ O-Cu-O bonds) and 
next nearest neighbor O-ions (linked by O-Cu-O bonds at 180$^o$), respectively. 
The average oxygen charge density $\rho$ is $-(1.5 + \epsilon(x))$, where $\epsilon(x)$  can be both positive or negative and is small for the
underdoped and optimally doped materials. 
The true charge on the Cu-ions may be slightly larger than 1, and can also be weakly doping dependent (see section ~\ref{explanations}).  
The charge carriers are the same in the optimally electron-doped cuprates (holes on the $\rho=0.5$ O$^{1-}$ sites) as in the hole carriers,  
which explains the hole-like transport behavior at low temperatures in the former \cite{Dagan07a}.
 
We do not present calculations to prove the valence transition.
The preponderance of parameters that are explicitly or implicitly included in Eq.~\ref{ionicity} implies that such a transition can always
be found from calculations, with the problem reduced to a numerical exercise. 
Whether or not the valence transition is actually occurring can only be determined by comparing against the experimental puzzles described
in section~\ref{challenges}. This is what is done in the next section. 

\section{Experimental ramifications of the valence transition model}
\label{explanations}

Fig.~3 shows a schematic phase diagram for both the conventionally prepared $T^{\prime}$ electron-doped cuprates and the hole-doped materials.
We show that {\it all} the experimental
features that are difficult to understand within the traditional models, as well as features thought to have contradictory interpretations,
have straightforward simple explanations within the valence transition model.
\begin{figure}
\includegraphics[width=3.2in]{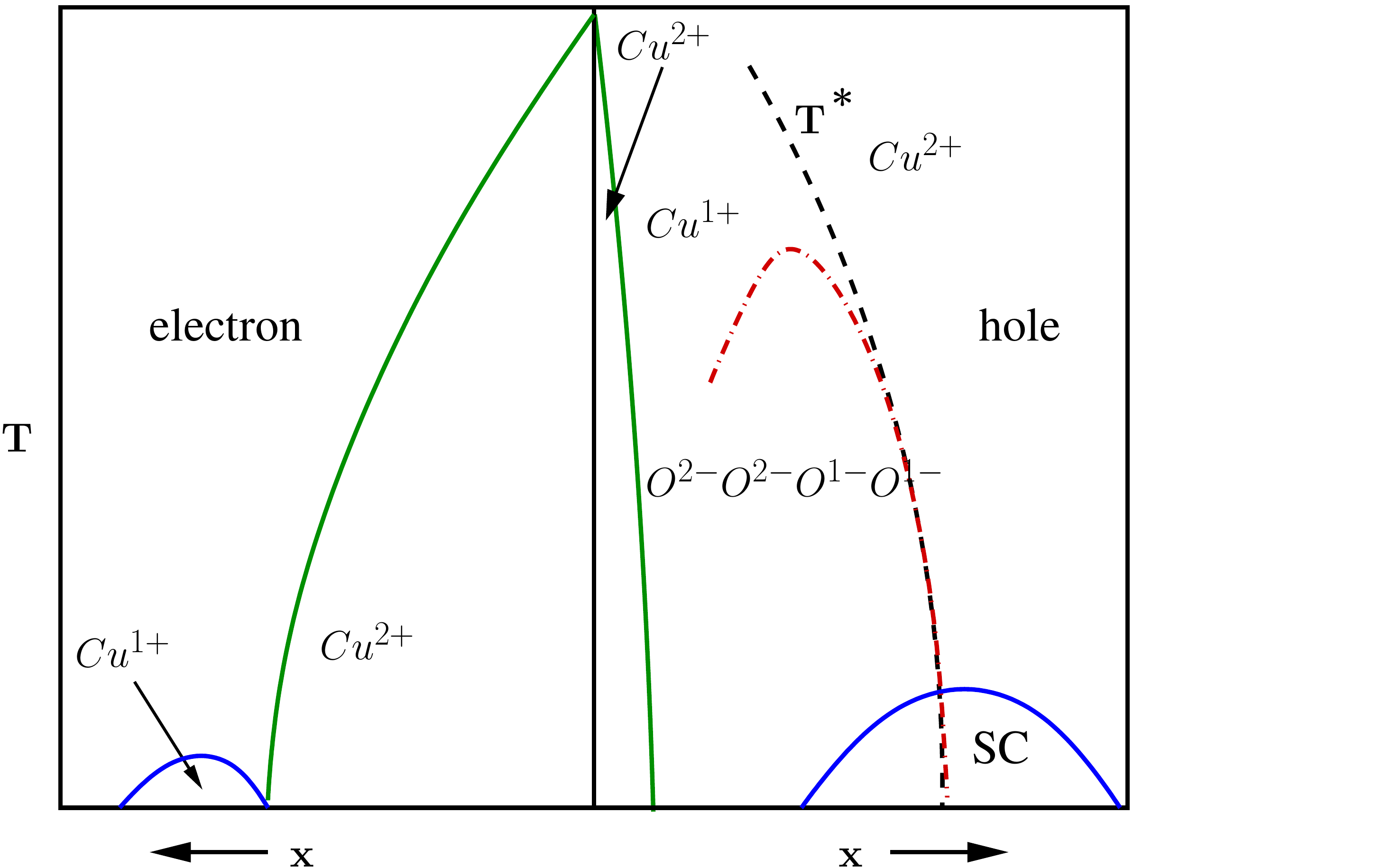}
\caption{Schematic phase diagram of the electron- and hole-doped superconducting cuprates within the valence transition model. 
The CO region has charge modulation O$^{2-}$-O$^{2-}$-O$^{1-}$-O$^{1-}$ along any Cu-O bond direction. The precise quantum critical point
at which T$^*$ intersects the dopant axis in the hole-doped systems cannot be evaluated without detailed calculations, and is not relevant.}
\label{Phase-diagram}
\end{figure}

\subsection{Electron-doped materials}
\label{Explanation-electrons}

The peculiarities observed with the electron-doped materials are all manifestations of the reduced $\Delta E(M,n)$ in the doped or
oxygen-deficient materials, and of the valence transition.
\vskip 0.5pc
\noindent \underbar{(i) Robust AFM and absence of coexisting SC and AFM}
\noindent \underbar{ in conventional $T{^\prime}$ compounds.}
Robust AFM has already been explained in the above: it is due to pinning of Cu$^{1+}$ to Ce$^{4+}$. We argue below
that SC emerges from the effective $\rho=0.5$ O-band, in which case the dramatic change in the electronic structure at optimal
doping \cite{Abe89a} as well as absence of coexisting SC and AFM \cite{Motoyama07a,Saadaoui15a} 
are not only expected, they are requirements.
\vskip 0.5pc
\noindent \underbar{(ii) Size of RE ion and AFM-SC boundary.} The larger the ionic radius, the smaller is the Madelung
energy stabilization of higher charge. The loss of AFM and appearance of SC at smaller doping with larger rare earth ions
\cite{Naito02a,Saadaoui15a} is therefore due to smaller $\Delta E(M,n)$ in these cases.
\vskip 0.5pc
\noindent \underbar{(iii) Oxygen deficiency as a requirement for SC.}
Reduced oxygen content reduces the absolute value of $\Delta E(M,n)$ in Eq.~\ref{ionicity} severely, because of
charge imbalance. It is even likely that it is this
reduced oxygen content that brings the system close to the boundary of the inequality Eq.~\ref{ionicity} in the first place. 
The peculiar dependence of the superconducting versus nonsuperconducting behavior on the oxygen partial pressure $p_{O_2}$, with the
superconducting materials lying in the stability region corresponding to Cu$_2$O \cite{Kim93a,Navarro01a} in which the Cu-ions are monovalent 
is thus a strong confirmation of the valence transition model. 
\vskip 0.5pc
\noindent \underbar{(iv) SC in the undoped $T^{\prime}$ thin films.} This is simply a consequence of 
smaller $\Delta E(M,n)$ in thin films with weak 3D contribution to the Madelung energy, such that the configuration with Cu$^{1+}$ is lower in energy even without doping.
Highest T$_c$ at zero doping is a signature that the paired Wigner crystal  
which is a precursor to SC, is most stable \cite{Clay18a,Li10a} at exactly $\rho=0.5$
or very close to this filling. This is where the superconducting pair-pair correlations are the strongest \cite{Gomes16a,DeSilva16a} (see below).
\vskip 0.5pc
\noindent \underbar{(v) Sign of the charge carrier.} Since in both hole-doped cuprates and the optimally electron-doped cuprates
conductivity involves the same process O$^{2-}$-Cu$^{1+}$-O$^{1-} \to$ O$^{1-}$-Cu$^{1+}$-O$^{2-}$ (and paired motion of O$^{1-}$-O$^{1-}$ 
spin singlet in the superconducting state, see below) the same sign of charge carrier is to be expected.
\vskip 0.5pc
\noindent \underbar{(vi) Reduction in $^{63,65}$Cu NMR frequerncy and wipeout}
\noindent \underbar{of Cu NQR intensity.} These are consequences of the
Cu$^{2+} \to$ Cu$^{1+}$ valence transition. Tiny EFG \cite{Abe89a,Williams07a,Wu14a} is a natural consequence of the spherically symmetric 3d$^{10}$ configuration of Cu$^{1+}$, and should be common to both electron and hole-doped materials (see below).
\vskip 1pc
\noindent \underbar {(vii) Charge-order.} The electron-doped materials beyond the AFM region, and the hole-doped materials within the pseudogap phase
both consist of nearly $\rho=0.5$ O-band within the valence transition theory. {\it Hence the same period 4 CO (see section~\ref{challenges}) 
in both cases are expected.} We discuss the charge and bond modulations in detail in the next subsection.
\vskip 0.5pc
\noindent \underbar {(ix) Low RE solubility limit.} Each introduction of Ce$^{4+}$ ion in the $T^{\prime}$ compounds necessitates the
conversion of a Cu$^{2+}$ to a Cu$^{1+}$. Since following the valence transition all Cu-ions are already in the state Cu$^{1+}$, further
substitution of Nd$^{3+}$ by Ce$^{4+}$ becomes impossible. This is not true in the hole-doped materials, where each dopant bivalent cation converts
an O$^{2-}$ to O$^{1-}$, and there exist an abundance of O$^{2-}$ even following the valence transition.
\vskip 0.5pc
\noindent \underbar {(vi) SC and Zn-substitution effect.} As with the CO, SC in both electron and hole-doped cuprates emerges from the
same nearly $\frac{1}{4}$-filled O-band within Eq.~\ref{hamuv} within the valence transition model. Hence similar Zn-substitition effects are also to
be expected. The deleterious effect of Zn-doping will be discussed in detail in the next subsection. 

\subsection{Hole-doped materials}

We now discuss the perplexing experiments of section~\ref{challenges} in the hole-doped cuprates in view of 
the schematic phase diagram of Fig.~\ref{Phase-diagram}.
\vskip 0.5pc
\noindent \underbar{(i) NMR, NQR and Nernst measurements: one versus} 
\noindent \underbar{two-component description.} Valence transition at the PG boundary is behind the dramatic change
in the Cu ion spin lattice relaxation rate \cite{Warren89a} and magnetic susceptibility \cite{Johnston89a,Alloul89a} at T$^*$. The drop in Cu-spin susceptibility is {\it not} due to
pairing of Cu$^{2+}$-spins, but due to transition to the spinless Cu$^{1+}$. This viewpoint gives the simplest explanation 
of the wipeout of Cu-NQR intensity accompanying stripe formation in La-compounds \cite{Hunt99a,Singer99a}: the wipeout is not due to
disorder (which is not expected to show such dramatic effect anyway) but simply due to the tiny EFG expected with
spatially symmetric 3d$^{10}$ electronic configuration. {\it Importantly, this explanation simultaneously suffices also for the small EFG of the Cu-ions
in the optimally electron-doped materials \cite{Abe89a,Wu14a} but not in the parent semiconductors.} 

A microscopic picture for the two spin component model of 
Haase {\it et al.} \cite{Haase08a,Haase09a,Haase12a} and Barzykin and Pines \cite{Barzykin09a} emerges now:
for T $>$ T$^*$ the spins are predominantly on the Cu$^{2+}$ sites, while for T $<$ T$^*$ they are predominantly on the O$^{1-}$. The PG phase
transition is indeed {\it not} due to pairing, in agreement with the conclusions of Cyr-Choiniere {\it et al.} \cite{Cyr-Choiniere18a}. 
Yet preformed O$^{1-}$-O$^{1-}$ spin singlet will also occur in the Cooper pair density wave, as we discuss below. The actual difference between 
T$^*$ and T$_{CO}$ is very likely material dependent. 
\vskip 0.5pc
\noindent \underbar{(ii) Commensurate period 4 CO, ARPES, broken C$_4$}
\noindent \underbar{symmetry and IUC inequivalence of O-ions.} As discussed in section \ref{challenges} broken C$_4$ symmetry and a commensurate period 4 CO
coexist within the PG phase, and both are linked with IUC inequivalence between the O-ions. We present here what is probably the 
simplest explanation of these observations. An explanation of the polar Kerr effect \cite{Xia08a,Karapetyan12a,Karapetyan14a,Lubashevsky14a} is obtained simultaneously.

Following the valence transition the possibility arises for the O-ions of the $\rho=0.5$ 2D O-band to become inequivalent. The O-ions
are located on the vertices of a 2D frustrated checkerboard lattice, since  the O-Cu-O bonds with bond angles of 90$^o$ and 180$^o$
are of the same strength in the absence of direct O-O hopping. Inclusion of direct O-O hopping will reduce the frustration, which however will still 
be strong. In several previous papers the present author and his colleagues have demonstrated the existence of 
a {\it Wigner crystal of spin-paired electrons} in the $\rho=0.5$ 2D frustrated  lattice \cite{Clay03a,Clay02a,Li10a,Dayal11a} within
the extended Hubbard model of Eq.~\ref{hamuv}.
We do not present additional calculations here. Rather, we briefly summarize the earlier results to show how the O-based period 4 CO 
of spin singlets in the CuO$_2$ layers emerges. 

\begin{figure}
\includegraphics[width=3.3in]{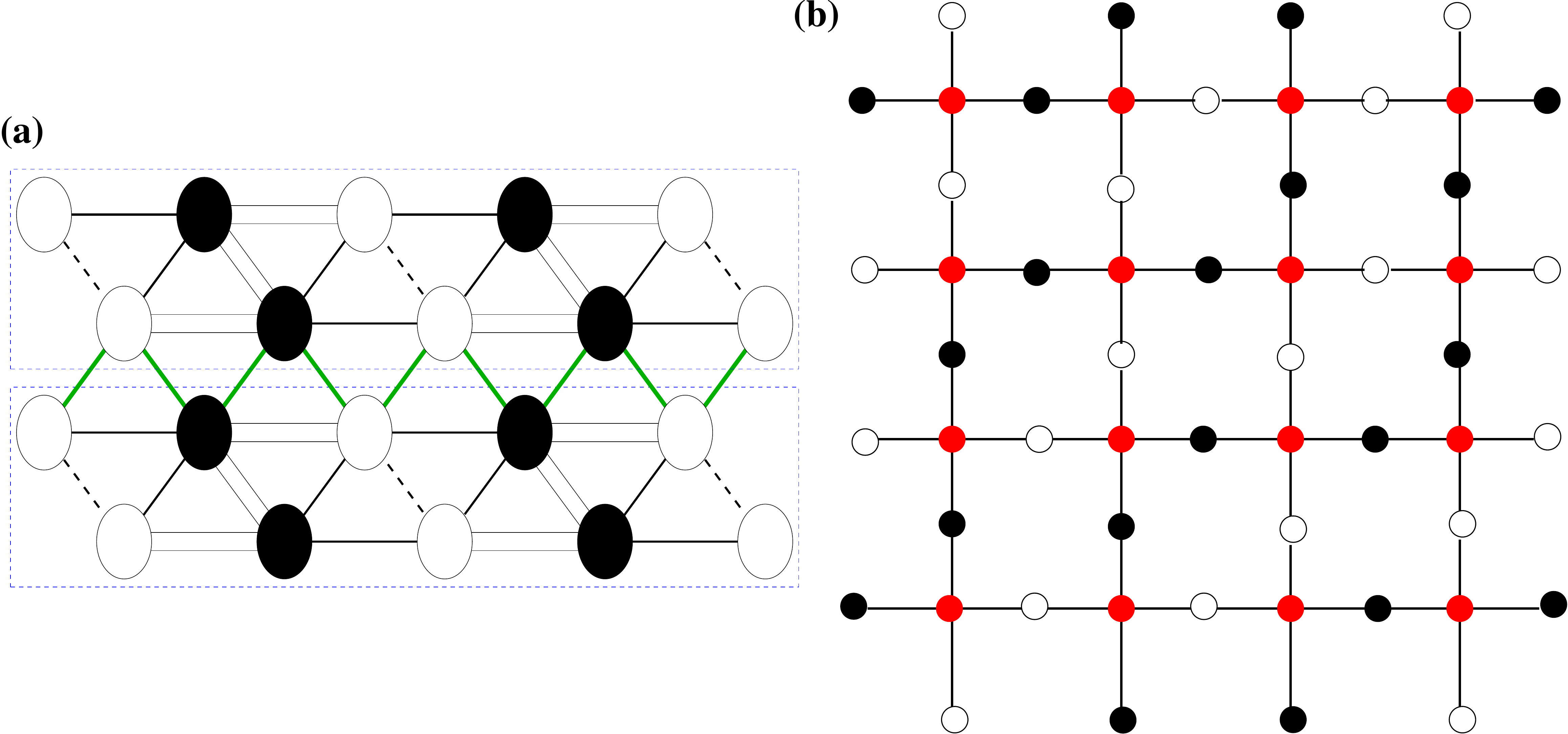}
\caption{(a) The paired Wigner crystal in the 2D $\rho=0.5$ anisotropic triangular lattice \cite{Dayal11a}.
Filled and unfilled circles correspond to charge-rich and charge-poor sites, respectively, while the 
dotted bonds are weaker than the bonds along the solid lines. Alternate
bonds are dimerized along the ``zigzag occupied horizontal stripes'' in the presence of lattice phonons \cite{Clay02a,Li10a,Dayal11a}, 
as in the purely 1D spin-Peierls dimerized chain.
The double bonds indicate spin-singlet pairing. (b) The paired Wigner crystal within the frustrated checkerboard $\frac{1}{4}$-filled O-sublattice in the
CuO$_2$ layer, consisting also of zigzag $\cdots$1100$\cdots$
charge occupancies, with spin singlet O$^{1-}$-Cu$^{1+}$-O$^{1-}$ bonds. The colors on the ions correspond to the same charges as in Fig.~\ref{ehdoped}. The CO consists of interpenetrating commensurate period 4 insulating ``stripes'' with O-hole occupancy $\cdots1100\cdots$
along the Cu-O bonds.}
\label{CO}
\end{figure}

We begin with the discussion of Eq.~\ref{hamuv} for the $\rho=0.5$ O$^{1-}$ holes on a monatomic 1D chain with only NN hopping and Coulomb interaction first 
(second neighbor electron hopping and V$^{NNN}$ are both zero). 
The ground state here is the simple Wigner crystal, $\cdots$1010$\cdots$,
where `1' and `0' denote charge-rich O$^{1-}$ and charge-poor O$^{2-}$  sites with actual charges 1.5 $\pm \delta$,
only for $V_p^{NN} > V_{pc}^{NN}$, where $V_{pc}^{NN} = 2|t_p|$ for $U_p \to \infty$ and is {\it larger} \cite{Clay03a} for finite $U_p$. 
For $V_p^{NN} < V_{pc}^{NN}$,
the charge-distribution is $\cdots$1100$\cdots$ even with V$^{NNN}=0$, driven by the strong tendency to form spin-singlet 
bond between the charge-rich 1--1 sites. 
This is the 2k$_F$ periodicity for the 1D lattice, and in the presence of
lattice phonons the CO is accompanied by bond distortion \cite{Clay03a}. Spin singlets are separated by pairs of vacancies, giving the
1D paired Wigner crystal. 
$V_{pc}^{NN}$ is smaller in the 2D square lattice, but even here the simplest  Wigner crystal charge occupancy is destabilized by geometric lattice frustration, driven by
by nonzero electron hopping along one or both diagonals of the square lattice. 
The ground state charge distribution now is the 2D paired Wigner crystal, 
with interpenetrating $\cdots$1100$\cdots$ COs along the two principal axes \cite{Li10a,Dayal11a}. 
While the paired Wigner crystal is again a quantum effect driven by tendency to form spin-singlets (for small $V_p^{NN}$) it is enhanced by $V_p^{NNN}$.
Nearest neighbor spin-singlet
coupled charge-rich pairs of sites are again separated by vacant pairs of sites.
The overall structure can also be thought of as alternating charge-rich and charge-poor {\it insulating stripes} as a result of this interpenetration. 
For illustration, we have shown in Fig.~\ref{CO}(a) the ``horizontal stripe'' structure demonstrated numerically for the anisotropic triangular lattice \cite{Clay02a}.
The occurrence of such paired Wigner crystal structures have been experimentally confirmed in a number
of $\rho=0.5$  2D organic charge-transfer solids, $\alpha$-, $\beta$- and $\theta$-(BEDT-TTF)$_2$X, some of which are superconducting under pressure \cite{Clay18a}.
\vskip 0.5pc
The tendency to the paired Wigner crystal in 2D is unique to $\rho=0.5$, as has been shown numerically \cite{Li10a,Dayal11a}. There
are multiple ways to understand this. First, only at this carrier density is such a paired CO commensurate, conferring it the exceptional stability that is necessary to 
dominate over both the metallic state as well as the single-particle Wigner crystal configuration. Alternately, as seen in Fig.~\ref{CO}(a), the 2D paired Wigner crystal
at $\frac{1}{4}$-filling consists of perfectly alternating exactly $\frac{1}{2}$-filled and exactly empty 1D chains,
which can occur only for $\rho=0.5$. The $\frac{1}{2}$-filled chain can be further stabilized by electron-phonon interactions that give the
spin-Peierls distortion.
The only requirement for this CO to occur at this density is
geometric lattice frustration \cite{Li10a,Dayal11a}. 
In Fig.~\ref{CO}(b) we have shown the charge occupancies of the paired Wigner crystal on the
O-lattice within the PG phase. The paired Wigner crystal structure for the checkerboard O-lattice is
arrived  at from our previous calculations \cite{Li10a,Dayal11a}, by simply insisting that the site occupancies are $\cdots$1100$\cdots$ along both principal axes.
The spin singlets consist of the 180$^o$ O$^{1-}$-Cu$^{1+}$-O$^{1-}$ bonds. 
In complete agreement with experiments
(see section~\ref{challenges}), the cuprate CO
is period 4 and oxygen-based. Below we point out that this charge occupancy will lead to bond distortions along the Cu-O bonds.

The strong O$^{1-}$-Cu$^{1+}$-O$^{1-}$ spin-singlet bonds along the Cu-O bond directions in Fig.~\ref{CO}(b) explain
the large antinodal gap that deviates from the simple $d$-wave form at the lowest temperatures in the underdoped samples, as seen in ARPES \cite{Hashimoto14a}. 
The CO in Fig.~\ref{CO}(b) is lacking in C$_4$ symmetry but possesses C$_2$ symmetry. The loss of C$_4$ symmetry is due to IUC inequivalence of O-ions, in agreement with
observations \cite{Lawler10a,Kohsaka12a}. The spins on the O$^{1-}$ are likely behind the weak magnetism observed in this state \cite{Lawler10a}.

\begin{figure}
\includegraphics[width=3.5in]{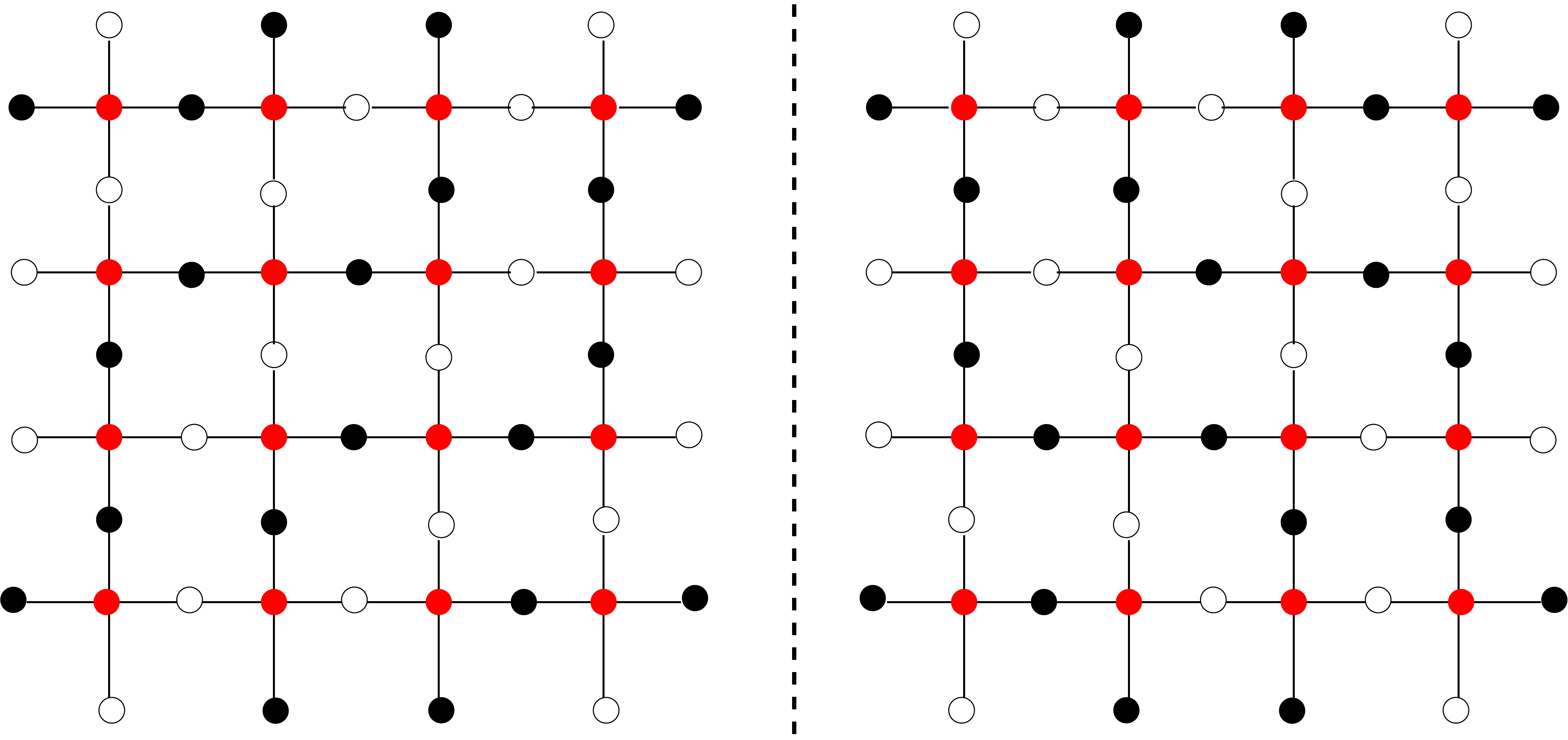}
\caption{Finite fragments of the CO structure related by reflection symmetry about the $y$-axis. From bottom to the top of the figures the zigzag 
effective filled chain goes from left to right in the left panel, and from right to left in the right panel. The mirror images cannot be superimposed
on one another. The same is true between the pair for reflection about the $x$-axis or either diagonal.} 
\label{chiral}
\end{figure}

We now address the polar Kerr effect \cite{Xia08a,Karapetyan12a,Karapetyan14a,Lubashevsky14a}.
As mentioned above, recent experimental work has shown that the Kerr rotation is not due to time reversal symmetry breaking, but to 
2D chirality \cite{Karapetyan12a}. The CO structure of Fig.~\ref{CO}(b) has broken reflection symmetries along both Cu-O bond directions (hereafter $\hat{x}$
and $\hat{y}$). There are two distinctly different diagonals along the $\hat{x}$ +  $\hat{y}$ direction and two other distinct
diagonals along $\hat{x} - \hat{y}$ (See Fig.~\ref{CO}(b)). Reflection symmetry along three of the four distinct diagonals are lost in the CO state of Fig.~\ref{CO}(b).
Only one of the four distinct diagonals of the infinite lattice continues to be a reflection plane. In any CO structure with 
{\it finite} domain size, however, 
it is entirely likely that this particular diagonal is nonexistent, in which case all reflection symmetries will be absent.  
Experiments indicate that the domain sizes in the CO phase 
\cite{Hoffman02a,Howald03a,Ghiringelli12a,Chang12a,Blackburn13a,Blanco-Canosa13a,Blanco-Canosa14a,Wu15a} are of size $\sim$ 20$a_0$. In Fig.~\ref{chiral} we have
shown schematics of the same CO structure of the Fig.~\ref{CO}(b) as well as its mirror image, which indeed cannot be superimposed on the original structure. One
prediction of this explanation of the polar Kerr effect is that the Kerr rotational angle should decrease with decreasing wavelength of the light, which ``sees''
smaller and smaller domains. The Kerr rotational angle in the experiments by Lubashevsky {\it et al.}, where the frequency of the light source is in the THz regime,
is in the milliradians \cite{Lubashevsky14a}. In contrast, the use of infrared light in the experiments by Karapetyan {\it et al.} gives rotation
in the microradians \cite{Karapetyan14a}. Additionally, the authors of reference \onlinecite{Lubashevsky14a} conducted their experiments as a function of
wavelength. Again, the rotational angle decreases with decreasing wavelength. 

In summary, within the valence transition model broken C$_4$ symmetry and polar Kerr rotation are indeed consequences of the O-based period 4 CO. 
\vskip 0.5pc
\noindent \underbar{(iii) Doping dependence of T$^*$} 
Experimentally, T$^*$ is largest in the highly underdoped systems (with the undoped system 
being AFM, however); with increasing doping, 
T$^*$ decreases  and finally vanishes at a critical doping on the dopant concentration axis. 
The reason for this doping dependence is understood qualitatively within the valence transition model. T$^*$ is primarily determined by decreasing
$\Delta E(m,n)$ and increasing $\Delta(W)$ within Eq.~\ref{ionicity}, as well as the stability of the commensurate period 4 CO
in Fig~\ref{CO}(b). The CO is most stable at exactly $\frac{1}{4}$-filling of the O-band and is gradually destabilized
away from that carrier concentration. The experimentally observed doping-independent CO commensurate periodicity 0.25$Q$ \cite{Mesaros16a,Cai16a} 
indicates that the valence transition occurs 
even for the weakest hole doping, while the lockin to this periodicity with doping suggests that aditional doped holes enter both the O-band (creating soliton-like
defects on the $\frac{1}{2}$-filled band 1D oxygen chains in Fig. \ref{Phase-diagram}) as well as on the Cu$^{1+}$ sites (the  
true charge 1.0 + $\epsilon(x)$ on the Cu-sites increases weakly with $x$). 
Decreasing T$^*$ with doping is then a manifestation of the shifting of the oxygen carrier density from commensurate $\rho=0.5$  
and of slight increase in Cu-ion charge
from the precise integer value of +1. With increaseed doping beyond the valence transition critical point there is weakening of  
the perfect order of Fig.~\ref{CO}, leading first to SC (see below) and finally the overdoped phase with $\rho$ significantly away from 0.5.
The quantum critical point on the dopant axis where the overdoped phase is reached cannot be determined from these qualitative observations and requires
actual calculations.

The above explanation of highest T$^*$ in the most underdoped hole materials suffices also for the highest superconducting 
T$_c$ in the undoped unconventionally prepared reduction annealed thin films of 
the electron-doped materials \cite{Naito16a}, since we argue that SC is a consequence of the destabilization of the paired Wigner crystal. 
It is likely that in the proposed phase diagram for the electron-doped materials in Fig. \ref{Phase-diagram}
the precise charge on the Cu-ions in the optimally doped systems is close to 1.15, such that the O-band is exactly $\rho=0.5$.
\vskip 0.5pc   
\noindent \underbar{(iv) Giant phonon anomaly.} As has been demonstrated numerically in our earlier work on monatomic $\rho=0.5$ systems
\cite{Clay03a,Li10a,Dayal11a} the paired Wigner crystal of Fig.~\ref{CO}(a) has a co-operative coexistence with a period
4 {\it bond-order wave} (BOW), whose order parameter is the expectation value of the nearest neighbor charge transfer, 
$(p^\dagger_{i,\sigma}p_{j,\sigma}+H.c.)$ in the present case. 
The latter, in turn, is coupled to the lattice phonons \cite{Clay03a,Li10a,Dayal11a}, making the
transition to the paired Wigner crystal a coupled charge-bond-lattice transition. Two possible period 4 BOWs can coexist with the $\cdots1100\cdots$
CO \cite{Clay17a}, with (i) the 1--1 singlet bond the strongest (S), the 1--0 bond of medium strength (M) and the 0--0 bond the weakest, giving an overall
bond modulation that is labeled $SMWM$; or with (ii) the 1--0 bond the strongest, the 1--1 bond weak and the 0--0 bond the weakest, giving
a bond modulation $SWSW^{\prime}$. Bond distortion (i) is a consequence of small to moderate Hubbard $U$, while the $SWSW^{\prime}$ pattern
occurs at large $U$. 

\begin{figure}
\includegraphics[width=3.0in]{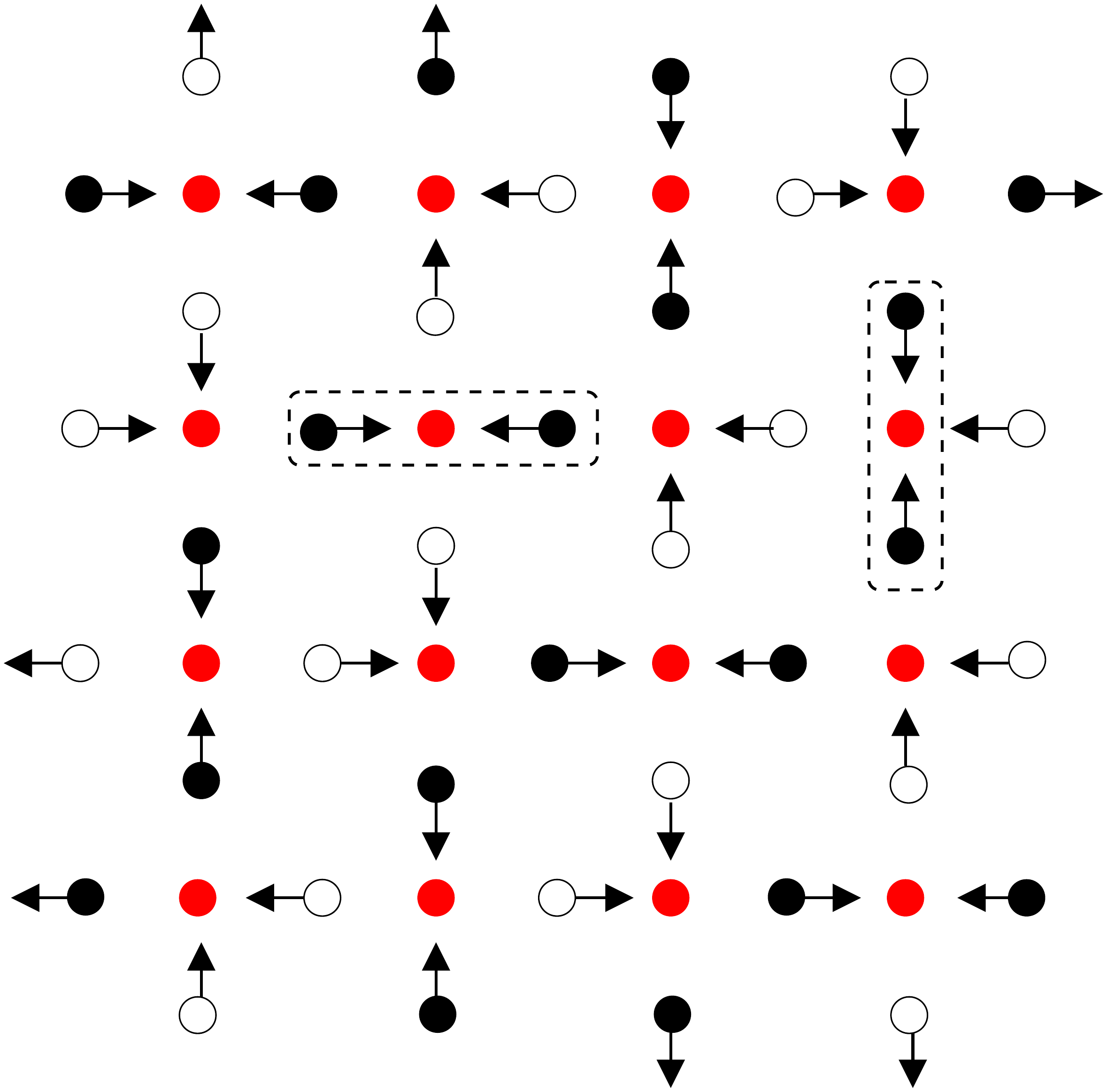}
\caption{The dominant phonon mode coupled to the paired Wigner crystal of the $\frac{1}{4}$-filled O-band. The dashed boxes indicate two of the
O$^{1-}$-Cu$^{1+}$-O$^{1-}$ spin singlets. The phonon mode in any one direction is the half-breathing mode of reference \onlinecite{Reznik06a}.}
\label{phonons}
\end{figure}

In Fig.~\ref{phonons} we have shown the dominant lattice phonon mode we expect for the two interpenetrating $SMWM$ bond distortion patterns
(see references \onlinecite{Clay03a,Li10a,Dayal11a}).
{\it We note that the lattice  distortion along any one direction is the same as the half-breathing mode (0.25, 0, 0) of Reznik 
et al.} \cite{Reznik06a,Reznik10a,Park14a}. 
On the other hand, the lattice distortion can also be thought of as spin-Peierls distortion of the ``zigzag'' $\frac{1}{2}$-filled chains consisting of 
alternating strong 180$^o$ spin-singlets O$^{1-}$-Cu$^{1+}$-O$^{1-}$ along the two Cu-O bond directions and weaker 90$^o$ O$^{1-}$-Cu$^{1+}$-O$^{1-}$ linkages.
We note that the dominant phonon mode in the electron-doped 
compounds were found to be (0.25, 0.25, 0) in early measurements \cite{Chen89a}.

\vskip 0.5pc
\noindent \underbar{(v) Preformed pairs.} Our paired Wigner crystal {\it is} the density wave of Cooper pairs  
proposed by other authors \cite{Anderson04b,Franz04a,Tesanovic04a,Chen04a,Vojta08a,Hamidian16a,Cai16a,Mesaros16a}. Earlier experimental
work had suggested such Cooper pairs along the Cu-O-Cu bond directions \cite{Kohsaka07a}. 
With weak doping that takes the system away from exact $\rho=0.5$ and there is increased lattice 
frustration, which ``melts'' the rigid CO giving the incoherent Cooper pairs seen in the Nernst effect measurements \cite{Wang05a,Wang06a,Li10b}.
{\it The preformed pair and competing broken symmetry scenarios for the PG phase are therefore not mutually exclusive; the different experiments
merely reflect the spin-paired nature of the CO.} 
\vskip 0.5pc
 \noindent \underbar{(vii) ARPES experiments.} The strong O$^{1-}$--Cu$^{1+}$--O$^{1-}$ spin-singlet bonds along the Cu-O bond directions in Fig.~\ref{CO}(b) explain 
the large antinodal gap that deviates from the simple $d$-wave form at the lowest temperatures in the underdoped samples. 
\vskip 0.5pc
\begin{figure}
\includegraphics[width=3.0in]{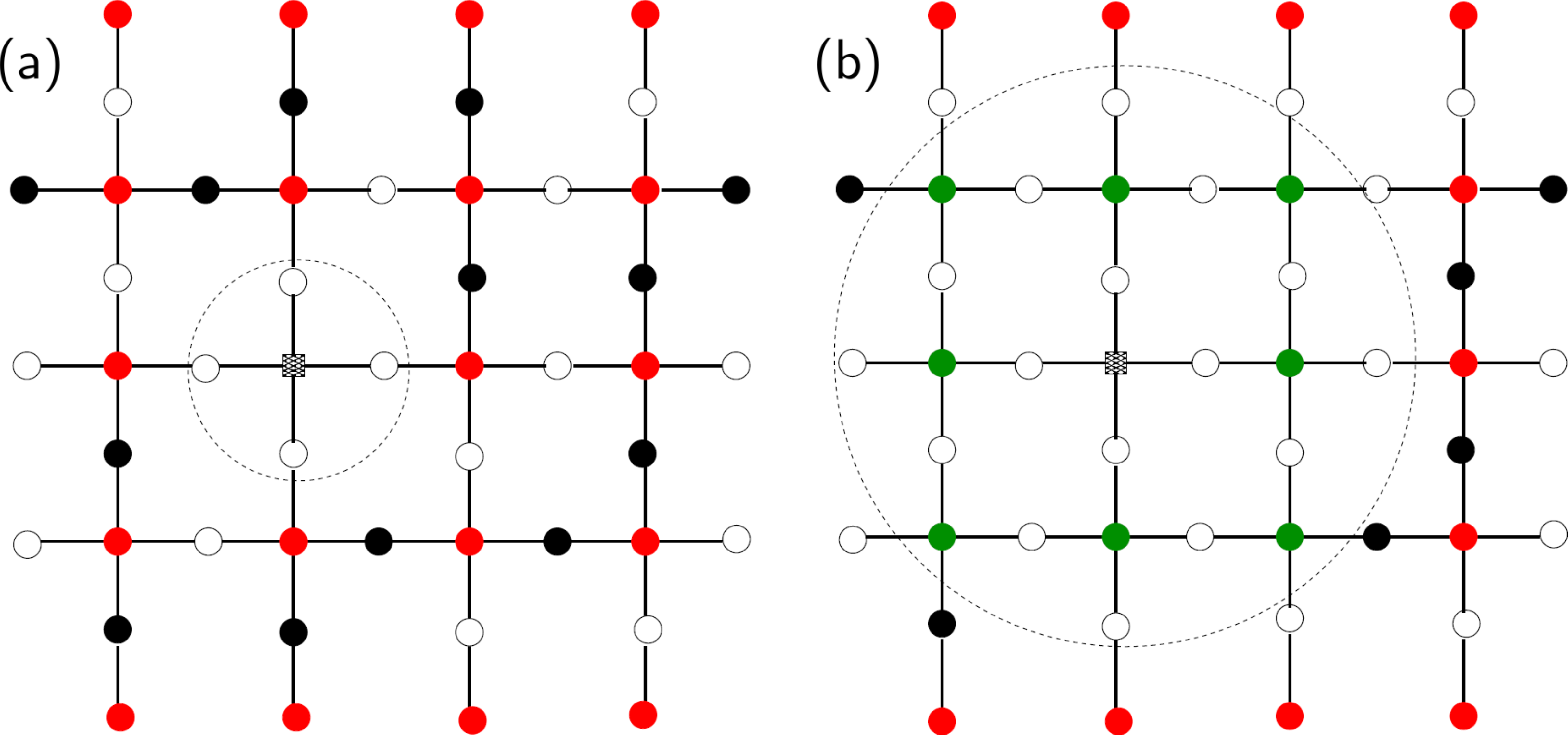}
\caption{Schematic of local reverse valence transition, Cu$^{1+} \to$ Cu$^{2+}$ upon Zn-doping. The different colors on the Cu and O-ions 
denote the same ionicities as in Fig.~2. (a) A single Zn$^{2+}$ ion (black square) replaces a Cu$^{1+}$
ion in the optimally doped electron-doped material or the hole-doped material in the pseudogapped or superconducting state, causing
O$^{1-} \to$ O$^{2-}$ transition in the immediate vicinity of Zn$^{2+}$. (b) Cu-ions that are neighbors of 
the first layer of newly formed O$^{2-}$ are now of charge +2, converting O$^{1-}$ further away from the Zn$^{2+}$ to O$^{2-}$, which in turn converts 
more distant Cu$^{1+}$ ions to Cu$^{2+}$. The cascading effect gives the charge-localized regions with spin moments on the Cu-ions. }
\label{Zn}
\end{figure}

 \noindent \underbar{(ix) Zn-doping and loss of SC, CO and PG.} The detrimental effect of Zn-doping on SC and the CO, in both electron- and hole-doped cuprates, 
as well as the 
semiconducting AFM nature of the Zn-doped materials, are due to the exceptional stability of the closed shell Zn$^{2+}$-ion, as seen in Fig.~\ref{IE}.
Within the valence transition model the Cu-ions are Cu$^{1+}$ in both the PG and superconducting states. In Fig~\ref{Zn} we show schematically
the consequences of Zn-doping. We imagine replacing a single Cu$^{1+}$ ion in the perfect CO state of Fig.~\ref{Zn}(a) with Zn, which necessarily
enters as Zn$^{2+}$. The immediate consequence is that the two neighboring oxygens which were previously singly charged O$^{1-}$ are now doubly charged
O$^{2-}$ (see  Fig.~\ref{Zn}(a)). The local gain in Madelung energy drives a reverse valence transition Cu$^{1+} \to$ Cu$^{2+}$ among the Cu-ions that
are neighbors of the newly formed O$^{2-}$ due to charge balance  requirments, 
as a consequence of which more distant O$^{1-}$ ions that are neighbors of second layer of  Cu$^{2+}$ have now higher charge O$^{2-}$ and so on, 
as is shown schematically in Fig.~\ref{Zn}(b). This ``cascading effect'' generates
phase separated charge localized regions with  spins on the Cu$^{2+}$ and superconducting regions with predominantly Cu$^{1+}$,
thus explaining the experimentally observed loss of the PG phase, as well as the so-called Swiss cheese 
model of reduced superfluid density. The Zn-driven transition is not due to the simple spinless character of Zn$^{2+}$.

\subsection{VB theory of correlated-electron bipolaronic SC.}
\label{VB-theory}
The key assumptions of the RVB theory of SC are that, (i) either the exactly $\frac{1}{2}$-filled band frustrated Mott-Hubbard semiconductor,
or the weakly doped system, is a quantum state that is a superposition of VB diagrams with NN singlet bonds \cite{Anderson73a,Fazekas74a};
and (ii) the NN singlets under appropriate condictions are mobile, and are hence configuration space equivalents of Cooper pairs \cite{Anderson87a}. 
The proposed spin-singlet state has not been found in any numerical investigation of 2D Hubbard models with realistic parameters. 
Yet the concept of mobile NN singlets being the equivalents of Cooper pairs
in configuration space has remained attractive. 
Many different propositions of valence bond solids (VBS)
and dimer liquids in the 2D $\frac{1}{2}$-filled band have therefore followed. To date, however, spin-singlet states at $\frac{1}{2}$-filling,
in systems with single orbital per site, 
have been found
within the Hubbard or Heisenberg Hamiltonian with realistic interaction parameters only in 1D chains and even-leg ladders. To the best of our knowledge,
the paired Wigner crystal that occurs in the frustrated $\rho=0.5$ \cite{Li10a,Dayal11a} is the only example of a spin-singlet state in 2D.

We have proposed a VB theory of correlated-electron SC \cite{Clay18a} wherein destabilization or ``melting'' of the paired Wigner crystal of Fig.~\ref{CO} 
by weak doping or increased frustration gives the $d_{x^2-y^2}$ SC in the cuprates. As shown in Fig.~\ref{phonons}, the pairing glue comes from
both the NN AFM spin-spin correlations as well as the lattice phonons. The SC within the proposed mechanism is a
coupled co-operative charge-spin-lattice effect. 
We cite recent numerical calculations by the 
author and his colleagues \cite{Gomes16a,DeSilva16a} to justify this theory. Additional justifications come from, (i) simultaneous
conceptual overlaps of the  present theory with the original RVB theory as well as the the original bipolaron theories of metal-insulator 
transitions and SC \cite{Chakraverty78a,Chakraverty79a,Chakraverty80b},
even as the formation of the bipolarons within the present theory is driven by AFM correlations and does not require overscreening of NN Coulomb repulsion;
and (ii) the preponderance of correlated-electron superconductors with $\frac{1}{4}$-filled bands (see Appendix). 

The necessary condition for SC driven by  electron-electron interactions is that the 
interactions {\it enhance} superconducting pair-pair correlations relative to the noninteracting
limit. As discussed in detail in section IIC, unbiased QMC and PIRG calculations within the weakly doped
2D Hubbard model have invariably found suppression of the superconducting pair-pair correlations with the Hubbard $U$. Very recently, similar calculations
of pair-pair correlations were performed for the first time in 2D frustrated lattices for the full range of bandfilling 0 to $\frac{1}{2}$ 
(carrier density $\rho$ per site 0 to 1) \cite{Gomes16a}. The quantity calculated was the average long-range ground state pair-pair correlation,
\begin{equation}
\bar{P} = N_P^{-1} \sum_{|\vec{r}|>2}P(r)
\end{equation} 
where $P(r) = \langle \Delta_i^\dagger \Delta_{i+r} \rangle$, 
$\Delta_i^\dagger$ is the pair creation operator of $d$-symmetry, and $N_P$ is the number of terms
in the sum \cite{Huang01a}. In obtaining the average pair-pair correlation only pairs separated by more than two lattice constants 
were considered, so that there was no
possibility of contamination from antiferromagnetic correlations. Computations were performed for
four different frustrated anisotropic periodic triangular lattices within
the Hubbard model, for Hubbard $U \leq 4|t|$. 
The computational techniques used were exact 
diagonalization for a 4$\times$4 lattice, PIRG for 6$\times$6 and 10$\times$6 lattices,
and constarined path quantum Monte Carlo (CPMC) \cite{Zhang95a} for the 10$\times$10 lattice. 
For each lattice, either the average $d_{x^2-y^2}$ or the $d_{xy}$ pair-pair correlation $\bar{P}$ was enhanced by Hubbard $U$ for a unique $\rho$ that was either exactly
0.5 or the density closest to this \cite{Gomes16a}. For all other $\rho$, including $\rho \sim 0.8-10$, the Hubbard $U$ suppresses pair-pair correlations. 
The calculations were then repeated using the temperature-dependent Determinantal Quantum Monte Carlo. For each lattice 
enhancement of pairing correlations was found uniquely for the same $\rho$ where the ground state calculations had found enhancement. 

More recently, similar calculations were performed \cite{DeSilva16a} for the
organic superconductors $\kappa$-(BEDT-TTF)$_2$Cu[N(CN)$_2$]Cl and $\kappa$-(BEDT-TTF)$_2$Cu$_2$(CN)$_3$. The calculations were for the {\it monomer} lattice 
of BEDT-TTF cations, for which $\bar{P}$ corresponding to $d_{x^2-y^2}$ pair-pair correlations were calculated for
two different periodic lattices (32 and 64 monomer molecules), with the electron hopping parameters for the two different compounds, again for all 
carrier densities $\rho$ per molecule. Once again, in every case
the Hubbard $U$ was found to enhance  $\bar{P}$ only for $\rho \simeq \frac{1}{2}$, and
suppressed the correlations for all other $\rho$. 

It is unlikely that the overall 
numerical results showing enhancement of superconducting pair-pair correlations uniquely at $\frac{1}{4}$-filling in eight different lattices, and
suppression at all other densities, is a coincidence. 
The logical conclusion that emerges is that exactly at $\frac{1}{4}$-filling the tendency to superconducting pairing is the strongest,
because of the unique stabilization of the paired Wigner crystal at this density.
Extending the above model to the checkerboard O-lattice in the CO state of the cuprates (Figs.~\ref{CO} and ~\ref{phonons}), 
the spin-singlets O$^{1-}$-Cu$^{1+}$-O$^{1-}$ constitute the Cooper pairs of the cuprate superconductors. Note that two recent 
theoretical calculations \cite{Khazraie18a,Wen18a} have suggested similar spin singlet formation in undoped BaBiO$_3$, and we comment on this
in the Appendix. 

The theory cannot be considered complete at the moment because the calculated pair-pair correlations do not exhibit
long range order (LRO) \cite{Gomes16a,DeSilva16a}. If superconducting long-range order is present at
finite $U$, $\bar{P}(U)$ would converge to a constant value as the
system size increases while $\bar{P}(U=0)$ would continue to
decrease. In this case $\bar{P}(U)/\bar{P}(U=0)$ would increase with
increasing system size.  In our results $\bar{P}(U)/\bar{P}(U=0)$ at
its peak value instead decreases with increasing system size. There are two possible reasons for this. The first is that true SC necessarily
requires additional interactions (for e.g., the electron-phonon interactions of Fig.~\ref{phonons}) ignored in the purely electronic Hamiltonian of
Eq ~\ref{hamuv}. Because explicit inclusion of electron-phonon interactions is required to realize the bond-distorted paired Wigner crystal
state \cite{Li10a,Dayal11a}, some role of electron-phonon interactions in the superconducting state might be expected. An alternate possibility
is that even as the current calculations indicate the likelihood of pair formation, the
question of LRO has to be settled by calculations of a correlation
function that is slightly different, because of the strong correlations between the pairs themselves within Eq.~\ref{hamuv}.
Elsewhere \cite{Mazumdar08a} we have attempted to simulate the paired Wigner crystal-to-SC
transition by performing exact diagonalization calculations on the periodic 4$\times$4  $\rho=1$ anisotropic triangular lattice
for $U<0, V>0$, with the assumption that the NN singlet bonds and pairs of vacancies in the paired Wigner crystal can
be thought of as double occupancies and single vacant sites, respectively. Transition from
a Wigner crystal of double occupancies to a $s$-wave superconductor occurs as the frustration is slowly increased \cite{Mazumdar08a}. Analysis of the
exact wavefunctions shows however that the only a subset of the many-electron configurations that describe the superconducting state
at $V=0$, $U<0$ dominate the $V>0$ wavefunction, giving partial support to the viewpoint that while a true superconducting state
is indeed reached $\rho=\frac{1}{2}$, more elaborate pairing correlations will be necessary to prove this.

\section{Pseudogap in S\MakeLowercase{r}$_2$I\MakeLowercase{r}O$_4$}
\label{iridates}
Sr$_2$IrO$_4$ has attracted strong attention in recent years as an effective square lattice Mott-Hubbard insulator with crystal structure similar to
that of the cuprates. The active layer consists of IrO$_2$ unit cells and the nominal charge on Ir in the compound is Ir$^{4+}$. With crystal field splitting this
gives the 5$d$ electron configuration as $t_{2g}^5$. 
The Mott-Hubbard gap now originates from the combined effects of spin-orbit coupling and repulsive Hubbard $U$. 
The $t_{2g}$ orbitals are split by spin-orbit coupling  into lower twofold degenerate total angular momentum J$_{eff}=\frac{3}{2}$ levels and an upper 
nondegenerate narrow J$_{eff}=\frac{1}{2}$ level \cite{Kim08a}. The $d^5$ occupancy of Ir$^{4+}$ then ensures single occupancy of the 
J$_{eff}=\frac{1}{2}$ level and Mott-Hubbard behavior.
AFM with N\'eel temperature comparable to that in the cuprates has confirmed this theoretical prediction.

Theoretical prediction of SC \cite{Wang11b} in electron-doped Sr$_2$IrO$_4$ has led to experimental studies that in turn 
indicate remarkable similarity between hole-doped cuprates and electron-doped Sr$_2$IrO$_4$ \cite{Yan15a,Torre15a,Kim16a,Battisti17a}.
The Mott-Hubbard gap vanishes abruptly at doping $\sim 5$\% and there emerges a ``nodal liquid'' with $d$-wave like gap near the nodal region, but once again, with 
strong deviation in the antinodal region where the gap is much larger. This pseudogap phase appears as {\it ``puddles'' of a phase-separated state around the
dopant atoms}, in regions where the doping is larger than a threshold value \cite{Battisti17a}. Very similar behavior was noted for the cuprates \cite{Kohsaka07a,Kohsaka08a}, making the mechanism of PG formation in doped Sr$_2$IrO$_4$ clearly
of interest. 

As with the hole-doped cuprates the valence transition model gives an easy to comprehend explanation of the PG. Furthermore, theoretical
predictions can be made here, as experimental studies have just began. Exactly as Cu$^{1+}$ 
has a very large ionization energy (Fig.~\ref{IE}) because of the closed shell nature of the ion, 
it is to be expected that the ionization energy of Ir$^{3+}$ with closed shell electron configuration $t_{2g}^6$ 
in the octahedral environment  is also very large. We believe that
the same valence transition from high to low charge, Ir$^{4+} \to$ Ir$^{3+}$ occurs here upon doping. This would either give the CO of Fig.~\ref{CO}
with O$^{1-}$--Ir$^{3+}$--O$^{1-}$ singlets, or a very strong tendency to this CO, which would explain the $d$-wave gap. 
We make two distinct theoretical predictions for electron-doped Sr$_2$IrO$_4$, which are both specific to the valence transition model and can be easily tested.

\section{Conclusion and experimental predictions} 
\label{conclusions}
The concept of negative charge transfer gap in hole-doped cuprates, conventionally electron-doped T$^\prime$ cuprates and
in undoped thin film T$^\prime$ compounds gives the simplest yet most comprehensive explanations for experiments which have been very difficult to understand
within the traditional models for cuprates. The VB theory of correlated-electron SC in the frustrated $\rho=0.5$ systems 
\cite{Gomes16a,DeSilva16a,Clay18a} is currently incomplete. However,
as we discuss in the Appendix, there exist a large number of superconductors that are or have been believed to be unconventional by many different
groups. In several cases, NN singlet pairing driven by electron-phonon interactions had been proposed \cite{Chakraverty79a,Chakraverty85a,Hague07a}.
Although this approach was severely criticized more recently \cite{Chakraverty98a}, we have shown that similar pairing can be driven also by 
electron-electron interactions \cite{Gomes16a,DeSilva16a}, provided the carrier density is exactly or close to $\rho=0.5$.
We point out below that the proponents of the bipolaron theory had missed the common carrier density 
$\rho=0.5$ that characterizes {\it all} the systems for which the original theory had been proposed. The present theoretical approach can thus provide a much needed ``global''
framework for understanding correlated-electron SC, where SC emerges from destabilization of realistic VBS that are very far from the
$\frac{1}{2}$-filled band limit.

We conclude by suggesting a series of experimental studies that can test the validity of the valence transition model.

(i) Copious amounts of O$^{1-}$ should occur in optimally electron-doped conventional T$^\prime$ cuprates as well as in the undoped superconducting thin films.
O$^{17}$ NMR measurements at optimal doping are suggested, in particular at high magnetic fields that have suppressed SC. Phonon anomalies similar to those in the hole-doped materials may be found in the CO states of electron-doped materials. The CO state should exhibit absence of C$_4$ symmetry.    

(ii) The deleterious effect of Zn-doping on SC and the CO states in electron-doped cuprates should be tested more carefully. The valence transition model predicts
destruction of SC by Zn-doping in underdoped PCCO and PLCCO. Such experiments are yet to be performed. It is not clear whether reduction annealed
thin films can be Zn-doped. But should this be possible, drastic reduction of superfluid density will be observed.

(iii) The consequences of Zn-doping on the PG state in the hole-doped cuprates should be tested more carefully than before. The bulk of the
experiments involving Zn-doping on the hole-doped materials have investigated the consequence on SC alone. The valence transition model predicts
equally strong deleterious effect on the PG.

(iv) O$^{17}$ NMR measurements are suggested for electron-doped Sr$_2$IrO$_4$. Within the traditional model of electron-doping, La-substitution of
the Sr$^{2+}$-ions merely generates Ir$^{3+}$ in an one-to-one fashion. Within the valence transition model for the PG phase transition, bulk
amounts of O$^{1-}$ are predicted in the PG state. We further predict IUC inequivalence of layer O-ions and broken C$_4$ symmetry in the PG state.

(v) As with the conventionally electron-doped cuprates, we predict low solubility of La-ions in Sr$_2$IrO$_4$. Once the PG state is reached, all the
Ir-ions are in the trivalent Ir$^{3+}$ state, and further doping becomes impossible.

(vii) Although the superoxygenated superconductor La$_2$CuO$_{4+\delta}$ has been known since the earliest
days, it remains much more poorly characterized than all other hole-doped systems. Within the valence transition model this material is just
the hole-counterpart of the oxygen-deficient undoped $T^\prime$ thin film compound \cite{Naito16a}. 
In both cases charge imbalance reduces $\Delta E_{M,n}$ much more
drastically than chemical substitution, driving the valence transition necessary for SC to occur. Indirect evidence for this is obtained from
the determination that a CO apears in this material with periodicity 0.25Q, without the LTO-to-LTT structural transition \cite{Zhang17a}. 
Once again, bulk amounts
of O$^{1-}$, loss of C$_4$ symmetry and wipeout of Cu-NQR intensity at the CO transition are predicted. Strongly deleterious effect of
Zn-substitution on the CO is predicted, should Zn-substitution be 
be possible while maintaining the superoxygenated character.

\section{Acknowledgments}

The author acknowledges close interactions and collaborations through the years with Professors David Campbell (Boston University) and 
R. Torsten Clay (Mississippi State University).
Much of the early work on the spatial broken symmetries in the $\frac{1}{4}$-filled band was done in collaboration with Professor Campbell.
Establishing the concept of the paired-electron crystal in the frustrated 2D $\frac{1}{4}$-filled band, the numerical demonstrations
of the enhancement of superconducting pair-pair correlations in such lattices, and the application of these concepts to organic 
charge-transfer solids would not have been possible without the continued collaboration with Professor Clay. The author
is grateful to Professor T. Saha-Dasgupta (Indian Association for the Cultivation of Science, Kolkata) 
for drawing his attention to the literature on negative charge-transfer gaps in nickelates and bismuthates.    
 
\section{Appendix}

Indirect support for the VB theory of correlated-electron SC presented in section ~\ref{VB-theory} is obtained by noting that there exist 
many different families of strongly correlated superconductors where the SC is limited to carrier concentration
exactly or close to $\frac{1}{4}$-filling. In many if not all cases SC is proximate to a broken symmetry state that 
is equivalent to the paired Wigner crystal. Although these materials have been of strong interest individually, only when they are 
considered together there emerges a pattern that suggests that SC is a generic feature of
strongly correlated $\frac{1}{4}$-filled band materials. The goal of this Appendix
is to point out this pattern. More extended discussions of organic charge-transfer solids can be found in reference 
\onlinecite{Clay18a}. 

\subsection{Superconducting (Ba,K)BiO$_3$}

Superconducting Ba$_{1-x}$K$_x$BiO$_3$ (T$_c \sim 30$ K) has been of strong interest also for three decades, while an earlier member of the
``family'' Ba(Pb,Bi)O$_3$ with T$_c = 15$ K has been known even longer. The first theoretical attempts to 
explain the SC here assumed that
the parent semiconductor BaBiO$_3$ contains charge-disproportionated Bi$^{3+}$ and Bi$^{5+}$ ions, creating a charge-density wave state that gave rise to
a gap at the Fermi surface.
SC was supposed to emerge from the doped charge-density wave, driven by coupling between electrons and breathing mode optical phonons \cite{Rice81a}.  
Systematic experimental investigations have failed to find this charge disproportionation \cite{Pei90a}. Equally importantly, estimates of electron-phonon couplings
based on DFT calculations \cite{Meregalli98a} were too weak to give T$_c \sim$ 30 K. Very recent computational studies that purportedly 
include the long-range Coulomb interactions have suggested that the actual electron-phonon couplings are much stronger \cite{Yin13a,Wen18a}. ARPES determination
of a Fermi surface much larger than what would be expected from the earlier DFT calculations is cited as evidence for the long-range Coulomb
interaction. However, similar larger-than-anticipated Fermi surface is also found in optimally electron-doped cuprates \cite{Horio16a}, as pointed out in 
section ~\ref{challenges}. 

An alternate approach to SC is suggested by recent theoretical and experimental work that have determined that
undoped BaBiO$_3$ is a negative charge-transfer gap material, with monovalent Bi ions, the Fermi level lying predominantly on O-based orbitals, and Bi 6$s$ orbitals significantly lower in energy \cite{Plumb16a,Khazraie18a}. It has been suggested that SC is due to hole pairs on oxygens, as had been suggested also in the
earlier version of this work \cite{Mazumdar89b}. In the following we present a slightly modified discussion of the mechanism of SC in view of the
present work.

The primary reason that the mechanisms of SC have been believed to be different for Ba$_{1-x}$K$_x$BiO$_3$  and cuprates is the 
difference that was thought to exist between the proximate semiconducting states in Ba$_{1-x}$K$_x$BiO$_3$  versus cuprates:
charge-density wave in the former, and AFM in the latter. The determination that CO is ubiquitous in the cuprates (see section~\ref{challenges})
makes this line of reasoning questionable. Similarly, the evidence for breathing mode like phonon coupling
in the cuprates \cite{Reznik10a}, also involving the O-ions, raises anew the question whether there exists a deep and fundamental relationships between the
perovskite oxides in general \cite{Mazumdar89b,Foyevtsova15a}. 

We believe that the need to go beyond existing theories arises from the following fundamental question, viz.,
why is SC in Ba$_{1-x}$K$_x$BiO$_3$ limited
to a relatively narrow dopant concentration \cite{Plumb16a,Pei90a}, $0.37 < x < 0.5$? 
The answer to this question is obtained by simple
counting of charge carriers. Given the homogenous charge of +3 on Bi, the reasonable assumption that charges on Ba (+2) and K (+1) are doping-independent  
confers average charge of $1.5-1.54$ to the O-ions, {\it exactly as in the cuprates}. But for the 3D nature of the bismuthates, we believe that the effective
electronic Hamiltonian that describes them and cuprates is the same, the $\rho=0.5$ O-based extended Hubbard Hamiltonian of Eq.~\ref{hamuv}.
As indicated in Fig.~\ref{phonons}, coupling to phonons resembling the breathing mode is a consequence of NN O$^{1-}$-pair to form 
spin singlets.
We continue below with discussions of other unconventional superconductors where also the carrier density of 0.5 (1.5) is the characteristic feature.

\subsection{Superconducting organic charge-transfer solids.} 
\label{CTS}
Superconducting charge-transfer solids (CTS) have the chemical formula M$_2$X or ZA$_2$, where X and Z are closed-shell inorganic 
anion and cation respectively, and M and A are organic molecules containing $\pi$-electrons. The charge density per molecule $\rho$ in the active organic layers is therefore exactly $\frac{1}{2}$.
SC in CTS can be proximate to AFM, spin liquid or CO, and is obtained from these exotic states by application of pressure instead of doping, 
{\it i.e.,} at constant carrier density $\rho$. In recent theoretical work, the present author and colleagues have shown that the CO in all cases
is a paired Wigner crystal \cite{Clay18a}, as evidenced either directly from the charge order pattern when it is known, or from a spin gap that is not expected from
competing charge order patterns. Direct calculations of superconducting pair-pair correlations \cite{DeSilva16a} for $0 \leq \rho \leq 1$ within the frustrated Hubbard model 
find enhancement of pair-pair correlations uniquely for $\rho \simeq \frac{1}{2}$. On the other hand, the observation that pressure-induced superconducting T$_c$ occurs 
at 4.8 K in $\kappa$-(BEDT-TTF)$_2$CF$_3$SO$_3$ while the ambient pressure N\'eel temperature is 2.5 K can be argued to indicate the inapplicability of spin-fluctuation theories.

\subsection{Superconducting cobalt oxide hydrate.} Layered cobaltates M$_x$CoO$_2$ (M = Li, Na, K) are strongly correlated-electron materials 
in which the carrier concentration can be varied over a wide range by varying the metal concentration $x$
\cite{Foo04a,Sakurai15a}. The electronically active components in these are CoO$_2$ layers separated by the M$^+$-ions. The Co-ions occupy an 
isotropic triangular lattice and have charge ranging from +3 (at $x=1$) to +4 (at $x=0$). 
The corresponding electron configurations are $t_{2g}^6$ with spin $S=0$, and $t_{2g}^5$ with $S=\frac{1}{2}$, respectively. 
Charge carriers are holes, with density per Co-ion $\rho=1-x$.  
Trigonal distortion 
splits the $t_{2g}$ orbitals into degenerate low-lying $e_g^\prime$ levels and a higher $a_{1g}$ level.
ARPES studies indicate that the $e_g^\prime$ levels are completely filled and therefore should be electronically inactive \cite{Hasan04a,Shimojima06a}, although this is somewhat controversial
\cite{Laverock07a}. Band calculations find a larger Fermi surface due to $a_{1g}$ levels, and a smaller Fermi surface due to the
$e_g^\prime$ levels \cite{Singh00a}. Thus the occupancy of the $a_{1g}$ orbital prinarily determines $\rho$ and the electronic behavior of the Co-ions in the
CoO$_2$ layer. The strongly $\rho$-dependent electronic behavior \cite{Foo04a} can only be understood within the triangular lattice nondegenerate
extended Hubbard model with moderate but finite Hubbard $U$ and nonzero NN interaction $V$ \cite{Li11a}.  

Superconducting Na$_x$CoO$_2 \cdot$ $y$H$_2$O ($x \sim 0.35$, $y \sim 1.3$) consists of the same CoO$_2$ layers, with the
H$_2$O entering in between the CoO$_2$ layers \cite{Takada03a}. 
A significant proportion of the water in the hydrated material
enters between the CoO$_2$ layers as H$_3$O$^+$ and the true superconducting composition 
is Na$_x$(H$_3$O)$_z$CoO$_2 \cdot$$y$H$_2$O \cite{Sakurai15a}.
Several chemical studies \cite{Sakurai06a,Banobre-Lopez09a,Sakurai15a} have found the Co-ion valency to be 
very close to +3.5. ARPES study \cite{Shimojima06a} puts the Co valence at +3.56 $\pm$ 0.05, making $\rho$ extremely close to 0.5. 
The successful theoretical modeling of the anhydrous material within the nondegenerate triangular lattice extended Hubbard model \cite{Li11a},
taken together with $\rho \simeq 0.5$ in the superconducting composition \cite{Sakurai06a,Banobre-Lopez09a,Sakurai15a,Shimojima06a} 
suggests strongly the applicability of the VB theory of SC described in
section~\ref{VB-theory}.  
 
\subsection{Superconducting spinels}
\label{spinelsc}

Spinels are inorganic ternary compounds AB$_2$X$_4$, with the
B-cations as the active sites. The B sublattice in the spinels forms
corner-sharing tetrahedra, giving rise to a geometrically frustrated
pyrochlore lattice. Out of several hundred spinel compounds with
transition metals as the B-cations only three undoped compounds are
confirmed superconductors, LiTi$_2$O$_4$ with T$_c \simeq 12$ K
\cite{Johnston73a}, CuRh$_2$S$_4$ (T$_c = 4.8$ K \cite{Hagino95a} and
up to 6.4 K under pressure \cite{Ito03a}), and CuRh$_2$Se$_4$ (T$_c =
3.5$ K \cite{Hagino95a}). Cu-ions in the latter compounds are monovalent \cite{Hart00a}, conferring
charge of +3.5 to Rh, which is the same charge on Ti in LiTi$_2$O$_4$. 
Band calculations have shown that the Fermi level in all cases lies in the t$_{2g}$ $d$-bands, and
are well separated from the empty e$_g$ bands 
as well as the completely filled $p$-bands due to O and S \cite{Satpathy87a,Massidda88a,Hart00a}.
The carrier densities in the
three spinel superconductors are then similar, with one $d$ electron per two Ti-ions in LiTi$_2$O$_4$,
and one $d$ hole per two t$_{2g}$ orbitals in CuRh$_2$S$_4$ and CuRh$_2$Se$_4$. This similarity
cannot be a coincidence, since simultaneously SC is absent in LiV$_2$O$_4$ \cite{Kondo97a}  
and CuV$_2$S$_4$ \cite{Seki92a,Hagino94a}, in which the carrier densities in the transition metal $d$ bands are 
only very slightly different.

Hint to the mechanism of spinel SC is reached by examination of the metal-insulator transitions in isostructural
isoelectronic CuIr$_2$S$_4$ \cite{Radaelli02a} and LiRh$_2$O$_4$ \cite{Okamoto08a}. In both cases the transitions are accompanied 
by $B^{4+}-B^{4+}-B^{3+}-B^{3+}$ ($B=$ Rh, Ir) charge and bond
tetramerization and $B^{4+}-B^{4+}$ spin-bonded dimers along specific
directions. This period 4 CO is exactly what is expected in the $\rho=0.5$ paired Wigner crystal \cite{Li10a,Dayal11a},
and can be understood within any theoretical model that lifts the degeneracy of the t$_{2g}$ orbital manifold and the
charge carrier occupies the nondegenerate $d$-band.
Both orbitally-induced band Jahn-Teller distortion \cite{Khomskii05a,Radaelli05a,Croft07a,Okamoto08a} and
spin-orbit coupling \cite{Rau16a,Kim08b,Witczak-Krempa14a} have been suggested as the drivers of the lifting of degeneracy
in CuIr$_2$S$_4$ and LiRh$_2$O$_4$. While spin-orbit coupling is weak in Ti, orbitally-induced band Jahn-Teller distortion,
especially in the presence of strong electron-electron interaction is conceptually feasible \cite{Clay10a}. SC in the
$\rho=0.5$ nondegenerate $d$-bands are then likely due to the motion of  Ti$^{3+}$--Ti$^{3+}$ and Rh$^{4+}$--Rh$^{4+}$
NN singlets. 

\subsection{Superconducting vanadium bronzes}
\label{V2O5}

Superconducting vanadium bronzes $\beta$-A$_{0.33}$V$_2$O$_5$, A = Li, Na, Ag,
have been of interest for as long as the charge transfer solids \cite{Chakraverty80a,Chakraverty80b}, and share
both $\frac{1}{4}$-filled band and pressure-induced CO-to-SC transition with the latter
(superconducting T$_c \simeq 6.5 - 8$ K at 8 GPa) \cite{Yamauchi02a,Yamauchi08a}.
The valence state of V$^{5+}$-ions in pure V$_2$O$_5$ is 3d$^0$. In $\beta$-A$_{0.33}$V$_2$O$_5$
there occur three different kinds of V chains and thus the
composition $\beta$-A$_{0.33}$V$_2$O$_5$ is stoichiometric. ARPES studies \cite{Okazaki04a} have shown that one of the three chains,
not known which, is $\frac{1}{4}$-filled, with exactly equal populations of V$^{5+}$ (3d$^0$) and
V$^{4+}$ (3d$^1$). There occurs a
dimensional crossover to quasi-2D behavior under pressure
\cite{Yamauchi08a}. The CO-to-SC transition is extremely sensitive to A-cation off-stoichiometry, with 
smallest non-stoichiometry destroying SC. This behavior is ubiquitous to all $\rho=0.5$ superconductors and is anticipated within
the VB theory of SC and our numerical calculations \cite{Gomes16a,DeSilva16a}. Interestingly, while vanadium bronzes were among the first compounds in which nearest
neighbor spin-paired bipolarons (V$^{4+}$-V$^{4+}$) were hypothesized
\cite{Chakraverty78a}, the role of the particular stoichiometric bandfilling was not noted by the investigators. All the materials
that were proclaimed to be bipolaronic insulators (paired Wigner crystal according to us) or bipolaronic superconductors
(Ti$_4$O$_7$, LiTi$_2$O$_4$, $\beta$-A$_{0.33}$V$_2$O$_5$) by these authors were $\frac{1}{4}$-filled.

\subsection{Superconducting Li$_{0.9}$Mo$_6$O$_{17}$}
\label{LMO}

Li$_{0.9}$Mo$_6$O$_{17}$ exhibits quasi-1D behavior at high temperature, a poorly understood metal-insulator transition
at $\sim$ 25 K \cite{Greenblatt88a}, and SC below 2 K \cite{Schlenker85a}. Large upper critical field for magnetic field parallel to the conducting
chains has led to the suggestion of triplet pairing \cite{Lebed13a},
although this has not been confirmed experimentally yet. The active electrons belong to the Mo $d$-orbitals. Two of the six Mo-ions
in the unit cell occur in tetrahedral sites; of the remaining four Mo-ions in octahedral sites, two form highly 1D two-leg zigzag ladders \cite{Merino12a}
or double zigzag chains \cite{daLuz11a}. DFT calculations \cite{Popovic06a} for LiMo$_6$O$_{17}$ have led to the interpretation that the true chemical formula should be
written as Li$^{1+}$(Mo$^{4.5+}$)$_2$Mo$^{\prime 6+}_4$(O$^{2-}$)$_{17}$ where Mo
but not Mo$^{\prime}$ constitute the coupled zigzag chains  \cite{Merino12a}. 
Mo-ion valence of +4.5 implies electron configurations of 4d$^0$ and
4d$^1$, while band structure calculations indicate that only the  d$_{xy}$
orbitals are the active bands, which are then exactly
$\frac{1}{4}$-filled. Giant Nernst effect \cite{Cohn12a} is yet another feature that Li$_{0.9}$Mo$_6$O$_{17}$
shares with the cuprates. 

\subsection{Superconducting intercalated and doped IrTe$_2$}
\label{IrTe2}

IrTe$_2$ consists of edge-sharing IrTe$_6$ octahedra with Ir layers sandwiched between Te layers
\cite{Ko15a}. The material is characterized by a poorly understood CDW transition at $\sim$ 260 K
that is accompanied by strong diamagnetism
and structural anomaly \cite{Oh13a,Fang13a}. The valence state \cite{Oh13a,Fang13a} of Ir at high temperatures is Ir$^{3+}$, with closed shell electron configuration 
t$_{2g}^6$. This would imply average ionic charge of -1.5 on the Te anions,
{\it i.e.} equal populations of Te$^{1-}$ and Te$^{2-}$. Thus the known cation and the anion valences here are exactly what we have proposed for doped
Sr$_2$IrO$_4$ in the above, with the carrier density $\rho=0.5$ in the Te-band (at least at high temperatures). 
One way to understand the diamagnetism following the CDW transition is
to assume NN Te$^{1-}$-Te$^{1-}$ spin singlet bonds, of the kind that occur in Ti$_4$O$_7$ and Na$_{0.33}$V$_2$O$_5$ (see reference \onlinecite{Chakraverty78a} and above).
There is, however, controversy as to whether the 260 K transition involves only
the Te ions, or both Ir and Te \cite{Fang13a}. SC appears upon intercalation of Pd into IrTe$_2$ (giving Pd$_x$IrTe$_2$) or in the Pd-doped
compound Ir$_{1-y}$Pd$_y$Te$_2$ (T$_c \sim$ 3K) for $x$ and $y$ larger than 0.02 and smaller than 0.1. This is yet another similarity with the materials
discussed here, viz., SC occurring over a very narrow carrier concentration range. CDW involving only the Te-ions, with period 4 
charge distribution Te$^{2-}$-Te$^{2-}$-Te$^{1-}$-Te$^{1-}$ would be expected within our theory. Additional experiments are necessary to determine
the precise natures of both the CDW and the SC here; the high temperature valences are certainly suggestive of a mechanism of SC common to all the 
materials discussed in the above. 

\subsection{Superconducting fullerides} Limitation of SC to a particular carrier concentration 
is a feature that superconducting fullerides
share with all correlated-electron superconductors discussed in the present work. 
Although complexes with molecular charges from -1 to -6 (including noninteger charges) are known \cite{Gunnarsson97a} , only those with
anioninc charge -3 are superconductors.
The observations of AFM in Cs$_3$C$_60$ with N\'eel temperature of 46 K, and pressure-induced AFM-to-SC transition at 38 K are both reminescent of the
widely noted behavior of superconducting $\kappa$-(BEDT-TTF)$_2$X (see above). A single spin per C$_{60}$ molecule is involved 
in the AFM \cite{Ganin10a,Klupp12a}. This has a unique explanation, viz., Jahn-Teller instability lifts the three-fold degeneracy of $t_{1u}$ MOS
of the trianion, with 2, 1 and 0 electrons occupying nondegenerate MOs with increasing energy. The system is now a Mott-Jahn-Teller insulator, with
the unpaired electron contributing to AFM. Existing theories of SC \cite{Gunnarsson97a,Capone09a} assume that pressure leads to the Jahn-Teller metal
that has regained the threefold degeneracy of the undoped material, and superconducting pairing is ``on-ball'', driven largely by electron-phonon 
coupling with the Hubbard $U$ playing either a competing or a co-operative role. The uniqueness of molecular charge -3 is not understood within these theories.

Within an alternate theoretical approach that fully explains the unique character of the trianion assumes that the loss of degeneracy is only partial. 
In a correlated-electron ion, the gain in energy due to Jahn-Teller instability is smaller when the orbital occupancy is by an even number of electrons
than when the occupancy is odd. This implies that in the AFM the energy gap between the doubly occupied and singly occupied antibonding MOs in the
Mott-Jahn-Teller insulator is smaller than that between the singly occupied and the vacant MO, with the difference between the gaps increasing with
the Hubbard $U$. Then in the correlated Jahn-Teller metal it is conceivable that the 
gap between the doubly occupied and singly occupied antibonding MOs is washed out by intermolecular hopping, 
which are nevertheless below the completely unoccupied MO due to
the same band Jahn-Teller denegeracy that characterizes spinel superconductors (see above). The molecular degeneracy is as in the 
spinels CuRh$_2$S$_4$ and  CuRh$_2$Se$_4$, with lower energy doubly degenerate MOs with electron populations of 1.5 electrons each, 
and a higher energy vacant MO \cite{Dutta14a}. Such on ``orbital reordering'' that takes the system from AFM to a singlet superconductor would be
similar to what has been found in calculations of pairing correlations \cite{DeSilva16a} for the $\kappa$-(BEDT-TTF)$_2$X. The pairing in this case
would be ``interball'' rather than ``intraball''. Experiments that can distinguish between the two kinds of pairing are needed to distinguish between the 
proposed theories.

\end{document}